# Observation of High-Order Anisotropic Magnetoresistance in a Cubic Ferromagnet

Haoran Chen[1, #], Yue Chen[2, 3, #], Yizi Feng[1], Ruda Guo[2, 3], Yuanfei Fan[1], Hongyue Xu[1], Tong Wu[1], Zhongxun Guo[1], Di Yue[1], Xiaofeng Jin[1], Yi Liu[6, 7], Zhe Yuan[3, †], Yizheng Wu[1,4,5, *]

[1]Department of Physics, Applied Surface Physics State Key Laboratory, Fudan University, Shanghai 200433, China

[2]The Center for Advanced Quantum Studies and School of Physics and Astronomy, Beijing Normal University, Beijing 100875, China

[3]Interdisciplinary Center for Theoretical Physics and Information Science, Fudan University, Shanghai 200433, China

[4]Shanghai Research Center for Quantum Sciences, Shanghai 201315, China

[5]Shanghai Key Laboratory of Metasurfaces for Light Manipulation, Fudan University, Shanghai 200433, China

[6]Institute for Quantum Science and Technology, Shanghai University, Shanghai 200444, China

[7]Department of Physics, Shanghai University, Shanghai 200444, China

## Abstract

High-order anisotropic magnetoresistance (AMR) is observed up to the 18th harmonic in cubic Fe(001) thin films, overturning the long-standing paradigm that only two- and four-fold terms are symmetry-allowed. Using angle-resolved transport and Fourier analysis, we show that six-fold and higher-order terms are intrinsic, tunable by temperature and thickness, and predicted by crystal symmetry. Microscopically, the two-fold sign reversal arises from a crossover between weak and strong scattering regimes, while high-order terms emerge from the interplay of anisotropic Fermi velocity and relaxation time. Our results establish high-order AMR as a symmetry-prescribed property of cubic ferromagnets, providing critical benchmarks for spin-orbit transport theory and enabling new angular-sensitive spintronic functionalities.

[#]These authors contributed equally.
[†]yuanz@fudan.edu.cn
[*]wuyizheng@fudan.edu.cn

Anisotropic magnetoresistance (AMR), discovered in 1856, remains a cornerstone phenomenon linking spin-orbit coupling to charge transport in ferromagnets [1-6]. Experimentally, AMR is typically characterized by rotating the in-plane magnetization with respect to the current direction. In polycrystalline metals, AMR shows a simple two-fold angular dependence [1,7,8], while in single crystals, reduced lattice symmetry leads to additional harmonics. In cubic ferromagnets such as Fe [9-18], Co [19,20], Ni [9,21-23], diluted magnetic semiconductors like (Ga,Mn)As [24,25], and ferromagnetic oxides [26-29], the prevailing paradigm, supported by decades of experiments [9-29], holds that the crystalline symmetry restricts AMR to two-fold and four-fold angular harmonics. Higher-order terms, such as six-fold components, have been considered forbidden in (001)-oriented films, and their observation remained exclusive to hexagonal or trigonal systems [30-36]. This widespread view has implicitly constrained the development of microscopic transport theories, which have largely focused on explaining only the lowest-order symmetries.

Yet, this long-standing assumption is not without tension. Early phenomenological symmetry analyses suggested that higher-order harmonics might be symmetry-allowed even in cubic crystals within a high-order expansion framework [1,21,37,38], but such high-order expansion was usually neglected due to the lack of experimental observation of high-order AMR in cubic magnets. Resolving this issue is critical, as the presence or absence of high-order AMR provides a definitive benchmark for microscopic models of spin-dependent transport, which must account for the complex interplay between anisotropic Fermi velocities and scattering processes [39-46].

Existing theoretical efforts at the microscopic level have primarily focused on the two-fold and four-fold AMR components. The anisotropic *s-d* scattering mechanism, originally proposed to account for resistivity anisotropy in ferromagnetic metals and alloys [1-4], provided a foundation for understanding AMR in disordered systems and was later extended to rationalize the sign of the two-fold AMR in cubic metals [1,5]. For single-crystal systems, well-defined band structures enable a more refined microscopic analysis, incorporating effects such as Fermi-surface topology [39], relaxation-time anisotropy [5,40-42], and Fermi-velocity anisotropy [43-46].

Nevertheless, a coherent microscopic theory including the high-order AMR components is still lacking. Even for the two-fold AMR in bcc Fe, which has been extensively studied in the past decades, the sign reversal with increasing temperature is still an open question [11-14].

In this Letter, we provide the first unambiguous experimental evidence of high-order AMR in epitaxial Fe(001) films—a prototypical cubic system with four-fold rotational symmetry. The high-order components exist with both in-plane and out-of-plane magnetization. Up to the 18th-order components out of our high-precision experiments overturn the prevailing paradigm that cubic ferromagnets host at most four-fold AMR. By integrating systematic experiments with first-principles calculations and a microscopic scattering model, we establish a comprehensive origin of these effects. The sign reversal of the two-fold AMR is shown to stem from a temperature-driven crossover between weak (momentum-conserving) and strong (phonon-mediated) scattering regimes, settling a decades-old puzzle in Fe. The six-fold and higher-order terms emerge naturally from the interplay between magnetization-dependent anisotropies in the Fermi velocity and the relaxation time. Our findings establish high-order AMR as an intrinsic and symmetry-prescribed property of cubic ferromagnets, providing critical experimental benchmarks for transport theory and opening new avenues for symmetry-engineered spintronics.

***Observation of high-order angular-dependent AMR***. —Epitaxial single-crystal Fe(001) films were grown on MgAl$_2$O$_4$(001) substrates by magnetron sputtering [18], where the minimal lattice mismatch ensures excellent crystalline quality [47]. Hall bar devices were patterned and measured in a superconducting vector magnet [Fig. 1(a)], which allows continuous in-plane rotation of the magnetic field defined by an angle $\phi_H$ relative to the current. Unless otherwise specified, all measurements were performed under $\mu_0 H = 1$ T, sufficient to saturate the magnetization.

At 300 K, the angular dependence of the longitudinal resistivity $\rho_{xx}(\phi_H)$ in a 5.8-nm-thick Fe film [Fig. 1(b)] is well described by the conventional two- and four-fold

AMR form, $\rho_{xx}(\phi_H) = \rho_0 + \Delta\rho_2 \cos 2\phi_H + \Delta\rho_4 \cos 4\phi_H$. In sharp contrast, at 5 K [Fig. 1(c)] the angular profile displays six maxima and minima within 360°, providing clear evidence of a $\cos 6\phi_H$ harmonic. To quantify these higher-order terms, we generalized the fitting function to an even-order cosine expansion,

$$\rho_{xx}(\phi_H) = \rho_0 + \sum_{n=1}^{\infty} \Delta\rho_{2n} \cos(2n\phi_H), \qquad (1)$$

and found that inclusion up to the eighth order ($n = 4$) yields satisfactory fits (see Supplemental Material [48] for details). Owing to the orthogonality of cosine functions, the extracted lower-order coefficients remain robust against truncation.

Since higher harmonics are unexpected in cubic (001) systems, we carefully examined possible extrinsic artifacts. A primary concern is a slight misalignment between magnetization $\boldsymbol{m}$ and applied field $\boldsymbol{H}$ due to magnetic anisotropy [18,20,52], which could mimic spurious higher-order terms. To test this possibility, we measured AMR under varying field strengths [Fig. 1(d)]. The $\Delta\rho_4$, $\Delta\rho_6$, and $\Delta\rho_8$ components evolve at low fields but saturate above 0.5 T, demonstrating that the six- and eight-fold terms observed at $\mu_0 H = 1$ T are intrinsic to the magnetization rotation. By contrast, the $\Delta\rho_2$ component grows linearly with field, consistent with ordinary magnetoresistance between $\boldsymbol{H} \parallel \boldsymbol{J}$ and $\boldsymbol{H} \perp \boldsymbol{J}$.

We measured the resistivity of a 4.6-nm-thick Fe(001) film by scanning both the azimuthal angle $\phi_H$ and the polar angle $\theta_H$ (see End Matter). The results reveal that high-order AMR harmonics are not confined to in-plane rotations but persist when the magnetization is rotated out of plane, extending across the full three-dimensional magnetization space.

*Up to 18th-order AMR revealed by Fourier analysis*. —The pronounced six-fold symmetry observed in Fig. 1(c) naturally raises the question of how high the harmonic order can extend in our system. Conventional fitting with truncated Fourier series cannot reliably separate higher-order contributions from noise. Terms beyond the eighth order add little improvement to the fit, making such approaches inconclusive. To enhance the signal-to-noise ratio (see Supplemental Material [48]) and directly resolve high-order angular harmonics, we measured the angular-dependent resistivity over ten consecutive rotation cycles [Figs. 2(a,b)] and analyzed the data using fast Fourier transformation (FFT) [Figs. 2(c,d)]. Thus, the resolution of the Fourier-transformed data is down to 1/10 fold, which is enough distinguish the high-order harmonics from a well-defined noise floor. Moreover, this procedure efficiently suppresses non-periodic noise and provides a reliable quantitative determination of high-order AMR components. Notably, distinct peaks are resolved up to the 18th order. And the amplitude of the six-fold component exceeds the two-fold component at 5 K [Fig. 2(c)], contrary to the common expectation of a monotonic decay with order.

The planar Hall effect (PHE) provides a complementary probe of the angle-dependent transverse resistivity $\rho_{xy}$ [53-55]. In cubic crystals, a reciprocal relation imposed by four-fold symmetry links AMR with ***J*** ∥ [100] to PHE with ***J*** ∥ [110], and vice versa [18,38,39]. As shown in Figs. 2(a-b), the reciprocity well established for two-fold terms also extends to higher-order harmonics, with AMR and PHE curves related by a 45° phase shift. The corresponding FFT spectra [Figs. 2(c-d)] further highlight striking symmetry selectivity. While the six-fold term dominates, the 4-, 8-, 12-, and 16-fold harmonics are strongly suppressed in PHE—a direct manifestation of Onsager reciprocity under four-fold symmetry [18]. Although such terms are forbidden by symmetry, weak residual amplitudes are observed, likely arising from substrate miscut or residual strain [18]. We also performed additional FFT analyses on a 78.6-nm Fe film, confirming that high-order AMR harmonics persist even in thicker samples (see Supplemental Material [48]), underscoring their robustness.

***Phenomenological theory of high-order AMR***. —FFT analysis establishes that angular-dependent AMR in Fe(001) films contains robust high-order harmonics, including the 6-, 10-, 14-, and even 18-fold terms. At first glance, the presence of these terms seems incompatible with the nominal four-fold crystalline symmetry of the (001) plane. To resolve this apparent contradiction, we invoke a phenomenological framework based on crystal symmetry [1,21,37], which systematically identifies the symmetry-allowed angular contributions.

In cubic (001) films, the conventional two- and four-fold AMR terms, as well as their reciprocal relation with PHE, have been thoroughly analyzed and experimentally confirmed [18,38,39]. Extending this framework, the in-plane resistivity tensor $\rho_{ij}$ can be expressed as a function of the magnetization angle $\phi_{mc}$ relative to [100],

$$\hat{\rho}(\phi_{mc}) = \begin{bmatrix} \rho_{11}(\phi_{mc}) & \rho_{12}(\phi_{mc}) \\ \rho_{21}(\phi_{mc}) & \rho_{22}(\phi_{mc}) \end{bmatrix}. \qquad (2)$$

Four-fold rotational symmetry imposes the constraints $\rho_{22}(\phi_{mc}) = \rho_{11}(\phi_{mc} + \pi/2)$ and $\rho_{21}(\phi_{mc}) = -\rho_{12}(\phi_{mc} + \pi/2)$, thereby reducing the number of independent functions to two. Since $\rho_{11}(\phi_{mc})$ and $\rho_{12}(\phi_{mc})$ are periodic, they can be expanded in Fourier series. Reflection symmetry and Onsager reciprocity further constrain the expansions: $\rho_{11}(\phi_{mc})$ contains only even-order cosine terms, whereas $\rho_{12}(\phi_{mc})$ is restricted to sine harmonics of the form $\sin[(4n-2)\phi_{mc}]$ with $n = 1,2,3,...$. By projecting onto the current directions [100] ($\phi_J = 0$) and [110] ($\phi_J = \pi/4$), with $\phi_{mc} = \phi_m + \phi_J$, we derive the AMR and PHE expressions [48]:

$$\begin{aligned} \rho_{xx}^{[100]}(\phi_m) &= \rho_0 + \sum_{n=1}^{\infty} \Delta\rho_{4n-2} \cos(4n-2)\phi_m + \sum_{n=1}^{\infty} \Delta\rho_{4n} \cos 4n\phi_m \\ \rho_{xy}^{[100]}(\phi_m) &= \sum_{n=1}^{\infty} \Delta\rho_{4n-2}^* \sin[(4n-2)\phi_m] \\ \rho_{xx}^{[110]}(\phi_m) &= \rho_0 + \sum_{n=1}^{\infty} (-1)^{n+1} \Delta\rho_{4n-2}^* \cos(4n-2)\phi_m + \sum_{n=1}^{\infty} (-1)^n \Delta\rho_{4n} \cos 4n\phi_m \\ \rho_{xy}^{[110]}(\phi_m) &= \sum_{n=1}^{\infty} (-1)^{n+1} \Delta\rho_{4n-2} \sin(4n-2)\phi_m \end{aligned} \qquad (3)$$

The amplitudes of the AMR harmonics for currents along [100] and [110] are governed by independent symmetry-allowed coefficients ($\Delta\rho_n$ and $\Delta\rho_n^*$), and their relative magnitudes are therefore not required to be identical. The symmetry analysis presented here is based on the $C_4$ point group, which already accounts for the tetragonal distortion of epitaxial Fe(001) films [56] while preserving four-fold rotational symmetry.

Our analysis shows that $\rho_{xx}^{[100]}$ and $\rho_{xy}^{[110]}$ share the same coefficient $\Delta\rho_{4n-2}$, whereas $\rho_{xx}^{[110]}$ and $\rho_{xy}^{[100]}$ share $\Delta\rho_{4n-2}^*$. Eq. (3) therefore reproduces the reciprocal relation between AMR and PHE in the high-order harmonics [48], and also reveals that even in cubic systems higher-order harmonics, such as $\cos 6\phi$ and $\cos 10\phi$, are symmetry-allowed. Thus, crystal symmetry not only permits but indeed prescribes the emergence of high-order angular terms in AMR and PHE, fully consistent with our experimental observations. Although such phenomenological expansions have long been recognized, our study provides the first direct experimental confirmation of high-order AMR in cubic (001) films.

***Temperature and thickness dependence of AMR.*** — The phenomenological analysis shows that six-fold and higher-order harmonics are symmetry-allowed, which naturally raises the question of under what conditions they emerge and become dominant. Figures 3(a) and 3(b) present representative AMR curves across a wide range of temperatures (5-300 K) and film thicknesses (3.8-97.7 nm), providing a systematic view of their evolution. At room temperature (RT), AMR is well described by two- and four-fold terms, whereas at low temperature a pronounced six-fold component clearly emerges.

All curves can be satisfactorily fitted using Eq. (1) up to the eighth order, with coefficients extracted relative to the magnetization angle to avoid misalignment artifacts (see Supplemental Material [48]). Figure 3(c) shows the temperature dependence of $\Delta\rho_2$, $\Delta\rho_4$, $\Delta\rho_6$, and $\Delta\rho_8$ in a 5.8-nm film. $\Delta\rho_2$ grows monotonically and reverses sign near 50 K. Both $\Delta\rho_6$ and $\Delta\rho_8$ vanish at high temperature, while $\Delta\rho_4$ exhibits a nonmonotonic evolution. The thickness dependence at 5 K is shown in Fig. 3(d). As thickness increases, $\Delta\rho_2$ changes sign from positive to negative near 5 nm, whereas $\Delta\rho_6$ undergoes a sign reversal around 20 nm. Notably, at the critical thickness where $\Delta\rho_2$ vanishes, $\Delta\rho_6$ remains large and positive, giving rise to the pronounced six-fold AMR in the 5.8-nm film [Fig. 1(c)]. In other regimes, the dominance of $\Delta\rho_2$ generally masks higher-order terms, rendering the six-fold component difficult to resolve. For quantitative comparison with previous literature, the absolute longitudinal resistivity $\rho_{xx}(T)$ for films of different thicknesses is shown in Fig. S11 of the Supplemental Material [48].

Figures 3(e-f) summarize these dependencies by mapping $\Delta\rho_2$ and $\Delta\rho_6$ as functions of temperature and thickness, effectively providing phase diagrams of their sign and amplitude. At high temperatures $\Delta\rho_2$ is positive for all thicknesses, but at low temperatures it becomes negative in thicker films, consistent with previous reports on Fe films [11-14]. This sign reversal has long been attributed to a competition between spin-orbit-induced AMR and the negative contribution from Lorentz-force-induced ordinary magnetoresistance (OMR) [13,14]. While field-dependent $\rho_{xx}$ measurements confirm that OMR yields a negative offset even at zero applied field, the sign reversal persists after subtracting this effect (see Supplemental Material [48]), proving that the intrinsic AMR itself changes sign. Figure 3(f) further reveals that $\Delta\rho_6$ is positive in thin films but negative in thicker ones at low temperatures, pointing to distinct film- and bulk-related mechanisms. With increasing temperature, $\Delta\rho_6$ in thick films decays rapidly, while in thin films it decreases more slowly. These systematic dependencies demonstrate that high-order AMR emerges most clearly when the two-fold term is suppressed, and further reveal distinct microscopic origins of the six-fold component in different thickness regimes.

*Microscopic mechanism of two-fold AMR sign reversal.* —The sign reversal of $\Delta\rho_2$ in Fe was reported decades ago [11-14], but its microscopic origin remains unresolved after excluding the OMR effect. A systematic interpretation of the sign of $\Delta\rho_2$ across different materials has only emerged recently [5,41] based on the extended *s-d* scattering model. The positive $\Delta\rho_2$ of bcc Fe at RT is attributed to two factors: (i) positive spin polarization of localized *d* states ($d_\uparrow > d_\downarrow$) at the Fermi level $E_F$, and (ii) the resistivity condition $\rho_\uparrow > \rho_\downarrow$, i.e. fewer spin-up *s* electrons at $E_F$. However, our calculated DOS from first principles shows a larger occupancy of the spin-up *s* electrons [Fig. 4(a)], consistent with $\rho_\downarrow > \rho_\uparrow$ calculated for bcc Fe [57]. Thus, our calculations suggest $\Delta\rho_2 < 0$ from *s-d* scattering, in agreement with low-temperature measurements along [100] and [110]. The observed sign reversal at higher temperatures therefore implies the involvement of additional competing scattering mechanisms.

Fully relativistic quantum-transport calculations that incorporate frozen thermal lattice disorder [58] reproduce the low-temperature negative $\Delta\rho_2$ and trace its continuous rise to positive values as phonon displacements grow [Fig. 4(b)]. Phonon scattering allows an electronic transition from $\boldsymbol{k}$ to $\boldsymbol{k'}$ with a finite momentum transfer, which was not included in the *s-d* scattering model [41,59]. We further applied a minimal scattering model to confirm the phonon contribution by computing the resistivity

$$\rho_{xx} = \sigma_{xx}^{-1} = \left[\frac{e^2}{V}\sum_{n\boldsymbol{k}} v_{n\boldsymbol{k}}^2 \tau_{n\boldsymbol{k}} \delta(\epsilon_{n\boldsymbol{k}} - E_\mathrm{F})\right]^{-1}, \tag{4}$$

where V is the real space volume, $v_{n\boldsymbol{k}}$ is the Bloch state velocity, and $E_\mathrm{F}$ is the Fermi energy. The transport relaxation time $\tau_{n\boldsymbol{k}}$ is evaluated from Fermi's golden rule, restricting the momentum transfer $|\boldsymbol{k}–\boldsymbol{k'}| \leqslant q$; see the inset of Fig. 4(c). At small $q$, corresponding to low temperatures, only long-wavelength acoustic phonon modes are excited. Most electron transitions occur with (nearly) conserved momentum, and the calculated $\Delta\rho_2$ is negative. [Fig. 4(c)]. As $q$ increases to ~1/20 of the Brillouin zone, $\Delta\rho_2$ turns positive after including electron scattering with finite $q$, consistent with the full quantum-transport calculation result. Although Fig. 4 presents result for current along [100], the same conclusions hold for current along [110] (see Supplemental Material [48]).

The thickness-dependent sign reversal in Fig. 3(e) originates from the competition between phonon and surface scattering. In thick films, phonon scattering dominates, and $\Delta\rho_2$ reverses its sign at a characteristic temperature where large-$q$ phonons are sufficiently excited. As film thickness decreases, temperature-independent large-$q$ processes induced by surface roughness scattering become more prominent, shifting the sign-reversal temperature lower. For ultrathin films ($d_\mathrm{Fe} < 5$ nm), surface scattering is dominant across all measured temperatures, rendering $\Delta\rho_2$ consistently positive. In the ultrathin limit, although quantum size effects may discretize the energy bands into quantum well states, these subbands largely preserve the orbital characters and scattering properties of their bulk counterparts [60,61]. Consequently, the observed sign change remains governed by the competition between scattering mechanisms rather than being a direct consequence of quantization. While quantum well states may introduce additional quantum oscillations in the AMR, such features are beyond the scope of this study.

***Microscopic picture of high-order AMR.*** —Although the competition between momentum-conserved and momentum-transfer scattering leads to the sign reversal of $\Delta\rho_2$, these mechanisms do not directly generate the higher-order terms of AMR. Instead, a general picture of high-order AMR can be elucidated from two sources [39,40,43]: the Fermi velocity and the relaxation time in Eq. (4), both of which depend on the magnetization direction modulated by spin-orbit coupling. An example of anisotropic Fermi velocity is shown in Supplemental Material [48]. Specifically in a cubic system, $\tau_{nk}$ and $v_{x,nk}^2$ can be expanded in terms of magnetization orientation $\phi$ as $c_0 + c_2 \cos 2\phi + c_4 \cos 4\phi + c_6 \cos 6\phi + \cdots$, i.e., $v_{x,nk}^2(\phi) = c_0^v + c_2^v \cos 2\phi + c_4^v \cos 4\phi + c_6^v \cos 6\phi + \cdots$, and $\tau_{nk}(\phi) = c_0^\tau + c_2^\tau \cos 2\phi + c_4^\tau \cos 4\phi + c_6^\tau \cos 6\phi + \cdots$. The products of $v_{x,nk}^2(\phi)$ and $\tau_{nk}(\phi)$ in the conductivity (and hence in the resistivity) naturally generate arbitrarily even-order harmonic terms $\cos 2n\phi$.

The two-fold component arises from the terms $c_0^\tau c_2^v$ and $c_0^v c_2^\tau$ summed over all $\mathbf{k}$ points. The four-fold component originates from three contributions: $c_0^\tau c_4^v$, $c_0^v c_4^\tau$, and $c_2^\tau c_2^v$. The nonmonotonic temperature dependence of $\Delta\rho_4$ likely reflects the competition among these terms. Analogously, the six-fold component arises from $c_0^\tau c_6^v$ and $c_0^v c_6^\tau$, as well as the cross terms $c_2^\tau c_4^v$ and $c_2^v c_4^\tau$. Consequently, the sign reversals of $\Delta\rho_6$ and $\Delta\rho_2$ are correlated, but they occur at different temperatures and thicknesses, as shown in Figs. 3(e) and 3(f). It can be rigorously proved that the 4$n$-fold terms in Eq. (3) are independent of current direction guaranteed by the in-plane fourfold rotational symmetry [48]. The (4$n$±2)-fold components, on the other hand, exhibit a two-fold angular dependence on the current direction in the (001) plane. These features fully agree with the experimental observations and the phenomenological theory.

*Conclusions*—We have demonstrated that epitaxial Fe(001) films host high-order AMR harmonics far beyond the conventional two- and four-fold terms. Fourier analysis reveals robust six-fold components and additional higher-order harmonics extending up to the 18th order. A phenomenological symmetry analysis confirms that such harmonics are symmetry-allowed in cubic (001) systems, while a microscopic picture based on the interplay of Fermi velocity and relaxation-time anisotropies naturally accounts for the high-order terms. A long-standing problem—the sign reversal of the two-fold AMR in Fe—is resolved: it evolves from negative values in the weak-scattering regime to positive values in the strong-scattering regime. Our results overturn the long-standing paradigm that cubic ferromagnets host only two- and four-fold AMR, establishing the existence of intrinsic high-order harmonics with implications for a broad class of cubic magnets. This work provides a unified microscopic framework linking symmetry, Fermi surface, and scattering to high-order AMR, thereby opening new opportunities for exploiting angle-dependent responses in spintronic devices.

## Acknowledgements

The work was supported by the National Key Research and Development Program of China (Grant No. 2024FYA1408500 and No. 2022YFA1403300), the National Natural Science Foundation of China (Grant No. 12274083, No. 12434003, No. 12221004, No. 12574115, and No. 12374101), the Shanghai Municipal Science and Technology Major Project (Grant No. 2019SHZDZX01), and the Shanghai Municipal Science and Technology Basic Research Project (Grant No. 22JC1400200 and No. 23dz2260100).

# End Matter

*Three-dimensional robustness of high-order AMR:*

To comprehensively establish that the high-order anisotropic magnetoresistance (AMR) reported in the main text is a general property and not only confined to the in-plane magnetization, we conducted further experiments to map its behavior throughout the full three-dimensional magnetization space. Demonstrating the persistence of this effect under out-of-plane magnetization is a critical step to confirm its origin in the material's intrinsic electronic band structure and spin-orbit coupling. For this purpose, we performed a full angular-dependent magnetotransport study on a 4.6-nm-thick Fe(001) film at 5 K, as shown in Fig. 5.

The experiment was carried out in a vector magnet system, which allowed for the continuous and independent scanning of both the azimuthal angle, $\phi_H$ (from 0° to 360°), and the polar angle, $\theta_H$ (from 0° to 90°), at a constant magnetic field strength, typically $\mu_0 H$=4 T. As depicted in Fig. 5(a), a polar angle of $\theta_H$=90° corresponds to the standard in-plane measurement configuration, while decreasing $\theta_H$ progressively tilts the magnetization vector out of the film plane. Owing to the large effective perpendicular anisotropy of the Fe film (~2.1 T), the magnetization polar angle $\theta_m$ deviates from the applied polar angle $\theta_H$. By contrast, the in-plane anisotropy field is much smaller (~0.05 T), making it a good approximation to assume that the magnetization azimuthal angle $\phi_m$ follows the applied field angle $\phi_H$. Figure 5(b) illustrates the evolution of the resistivity, $\rho_{xx}$, as a function of $\phi_H$ at several selected polar angles. As the magnetization tilts out-of-plane, the overall AMR amplitude diminishes due to projection effects. However, the complex angular features remain clearly evident, with the contributions from high-order harmonics still being pronounced at angles like $\theta_H$ = 60° and 75°. The complete three-dimensional resistivity map, $\rho_{xx}(\theta_H, \phi_H)$, shown in Fig. 5(c), visually confirms that these higher-order angular structures are smoothly preserved during the out-of-plane rotation, showing no abrupt changes or disappearance, thus proving their continuity and robustness in 3D space.

For a rigorous quantitative analysis, each $\rho_{xx}(\phi_H)$ curve at a fixed $\theta_H$ was fitted using Eq. (1) from the main text to extract the harmonic coefficients, $\Delta\rho_{2n}(\theta_H)$, up to the eighth order. To accurately analyze their behavior, the applied field angle $\theta_H$ was converted to the actual magnetization polar angle $\theta_m$ by accounting for the effective perpendicular anisotropy of the Fe film, see Supplemental Material [48]. The extracted coefficients are plotted as functions of $\theta_m$ in Figs. 5(d-g). All coefficients vanish as the magnetization becomes fully perpendicular to the film plane ($\theta_m=0°$), and their dependence can be excellently described by a series expansion of $\sin^{2n}\theta_m$ up to $n = 4$, consistent with theoretical expectations for the rotation of the resistivity tensor.

Crucially, the entire 3D mapping was repeated at various magnetic field strengths from 3 T to 6 T. The higher-order coefficients, $\Delta\rho_4$, $\Delta\rho_6$, and $\Delta\rho_8$, collapse onto a single curve, exhibiting no dependence on the applied field strength, as shown in Figs. 5(e-g). This provides irrefutable evidence that they are intrinsic effects originating from spin-orbit coupling. In stark contrast, the two-fold coefficient, $\Delta\rho_2$, shows a strong field dependence, clearly indicating that the OMR driven by the Lorentz force primarily influences this term.

To address the temperature dependence of the three-dimensional AMR, we additionally performed the same full angular-map measurements on the same 4.6-nm-thick Fe(001) film at room temperature (RT). The extracted coefficients $\Delta\rho_2$, $\Delta\rho_4$, and $\Delta\rho_6$ are shown in Fig. 5(h-j) as functions of $\theta_m$. While $\Delta\rho_2$ and $\Delta\rho_4$ remain finite, $\Delta\rho_6$ is nearly zero within the experimental uncertainty, demonstrating the absence of AMR harmonics higher than four-fold at RT. The red curves in Fig. 5(h, i) correspond to fits using a series expansion of $\sin^{2n}\theta_m$ truncated only at $n = 2$, indicating that no higher-order terms are required to describe the out-of-plane AMR at RT. The corresponding full angular map $\rho_{xx}$ ($\theta_H$, $\phi_H$) at RT, shown in Fig. 5(k), exhibits no discernible high-order angular features. These results demonstrate that high-order AMR harmonics emerge only at low temperatures.

In conclusion, these exhaustive out-of-plane measurements definitively confirm that the high-order AMR harmonics are not a phenomenon limited to two dimensions but are a robust, intrinsic property of the material that persists throughout the entire three-dimensional magnetization space.

# References

[1] T. McGuire and R. Potter, Anisotropic magnetoresistance in ferromagnetic 3d alloys. IEEE Trans. Magn. **11**, 1018 (1975).

[2] J. Smit, Magnetoresistance of ferromagnetic metals and alloys at low temperatures. Physica **17**, 612 (1951).

[3] L. Berger, Influence of spin-orbit interaction on the transport processes in ferromagnetic nickel alloys, in the presence of a degeneracy of the 3d band. J. Appl. Phys. **34**, 1360 (1963).

[4] I. A. Campbell, A. Fert, and O. Jaoul, The spontaneous resistivity anisotropy in Ni-based alloys. J. Phys. C: Solid State Phys. **3**, S95 (1970).

[5] S. Kokado and M. Tsunoda, Anisotropic magnetoresistance effect: general expression of AMR ratio and intuitive explanation for sign of AMR ratio. Adv. Mater. Res. **750-752**, 978 (2013).

[6] P. Ritzinger and K. Výborný, Anisotropic magnetoresistance: materials, models and applications. R. Soc. Open Sci. **10**, 230564 (2023).

[7] A. Kobs, S. Heße, W. Kreuzpaintner, G. Winkler, D. Lott, P. Weinberger, A. Schreyer, and H. P. Oepen, Anisotropic interface magnetoresistance in Pt/Co/Pt sandwiches. Phys. Rev. Lett. **106**, 217207 (2011).

[8] A. Philippi-Kobs, A. Farhadi, L. Matheis, D. Lott, A. Chuvilin, and H. P. Oepen, Impact of symmetry on anisotropic magnetoresistance in textured ferromagnetic thin films. Phys. Rev. Lett. **123**, 137201 (2019).

[9] D. C. Larson, J. E. Christopher, R. V. Coleman, and A. Isin, Magnetoresistance of nickel and iron single-crystal films. J. Vac. Sci. Technol. **6**, 670 (1969).

[10] M. Tondra, D. K. Lottis, K. T. Riggs, Y. Chen, E. D. Dahlberg, and G. A. Prinz, Thickness dependence of the anisotropic magnetoresistance in epitaxial iron films. J. Appl. Phys. **73**, 6393 (1993).

[11] U. Ruediger, J. Yu, S. Zhang, A. D. Kent, and S. S. P. Parkin, Negative domain wall contribution to the resistivity of microfabricated Fe wires. Phys. Rev. Lett. **80**, 5639 (1998).

[12] A. D. Kent, U. Rüdiger, J. Yu, L. Thomas, and S. S. P. Parkin, Magnetoresistance, micromagnetism, and domain wall effects in epitaxial Fe and Co structures with stripe domains (invited). J. Appl. Phys. **85**, 5243 (1999).

[13] P. Granberg, P. Isberg, T. Baier, B. Hjörvarsson, and P. Nordblad, Anisotropic behaviour of the magnetoresistance in single crystalline iron films. J. Magn. Magn. Mater. **195**, 1 (1999).

[14] R. P. van Gorkom, J. Caro, T. M. Klapwijk, and S. Radelaar, Temperature and angular dependence of the anisotropic magnetoresistance in epitaxial Fe films. Phys. Rev. B **63**, 134432 (2001).

[15] T. Hupfauer, A. Matos-Abiague, M. Gmitra, F. Schiller, J. Loher, D. Bougeard, C. H. Back, J. Fabian, and D. Weiss, Emergence of spin-orbit fields in magnetotransport of quasi-two-dimensional iron on gallium arsenide. Nat. Commun. **6**, 7374 (2015).

[16] M. W. Jia, J. X. Li, H. R. Chen, F. L. Zeng, X. Xiao, and Y. Z. Wu, Anomalous Hall magnetoresistance in single-crystal Fe(001) films. New J. Phys. **22**, 043014 (2020).

[17] Y. Miao, T. Li, X. Chen, C. Gao, and D. Xue, Temperature dependence of angular-dependent magnetoresistance in epitaxial Fe(001) film. J. Appl. Phys. **133**, 103902 (2023).

[18] H. Chen, Z. Cheng, Y. Feng, H. Xu, T. Wu, C. Chen, Y. Chen, Z. Yuan, and Y. Wu,

Anisotropic galvanomagnetic effects in single-crystal Fe(001) films elucidated by a phenomenological theory. Phys. Rev. B **111**, 014437 (2025).

[19] X. Xiao, J. H. Liang, B. L. Chen, J. X. Li, D. H. Ma, Z. Ding, and Y. Z. Wu, Current-direction dependence of the transport properties in single-crystalline face-centered-cubic cobalt films. J. Appl. Phys. **118**, 043908 (2015).

[20] Y. Miao, D. Yang, L. Jia, X. Li, S. Yang, C. Gao, and D. Xue, Magnetocrystalline anisotropy correlated negative anisotropic magnetoresistance in epitaxial $Fe_{30}Co_{70}$ thin films. Appl. Phys. Lett. **118**, 042404 (2021).

[21] W. Döring, Die abhängigkeit des widerstandes von nickelkristallen von der richtung der spontanen magnetisierung. Ann. Phys. **424**, 259 (1938).

[22] V. A. Marsocci, Magnetoresistance and Hall-voltage measurements on single-crystal Ni and Ni-Fe thin films. J. Appl. Phys. **35**, 774 (1964).

[23] X. Xiao, J. X. Li, Z. Ding, and Y. Z. Wu, Four-fold symmetric anisotropic magnetoresistance of single-crystalline Ni(001) film. J. Appl. Phys. **118**, 203905 (2015).

[24] A. W. Rushforth, K. Výborný, C. S. King, K. W. Edmonds, R. P. Campion, C. T. Foxon, J. Wunderlich, A. C. Irvine, P. Vašek, V. Novák, K. Olejník, J. Sinova, T. Jungwirth, and B. L. Gallagher, Anisotropic magnetoresistance components in (Ga,Mn)As. Phys. Rev. Lett. **99**, 147207 (2007).

[25] W. Limmer, M. Glunk, J. Daeubler, T. Hummel, W. Schoch, R. Sauer, C. Bihler, H. Huebl, M. S. Brandt, and S. T. B. Goennenwein, Angle-dependent magnetotransport in cubic and tetragonal ferromagnets: Application to (001)- and (113)A-oriented (Ga,Mn)As. Phys. Rev. B **74**, 205205 (2006).

[26] L. M. Wang and C.-C. Guo, Anisotropic magnetoresistance and spin polarization of $La_{0.7}Sr_{0.3}MnO_3/SrTiO_3$ superlattices. Appl. Phys. Lett. **87**, 172503 (2005).

[27] P. Perna, D. Maccariello, F. Ajejas, R. Guerrero, L. Méchin, S. Flament, J. Santamaria, R. Miranda, and J. Camarero, Engineering large anisotropic magnetoresistance in $La_{0.7}Sr_{0.3}MnO_3$ films at room temperature. Adv. Funct. Mater. **27**, 1700664 (2017).

[28] R. Ramos, S. K. Arora, and I. V. Shvets, Anomalous anisotropic magnetoresistance in epitaxial $Fe_3O_4$ thin films on MgO(001). Phys. Rev. B **78**, 214402 (2008).

[29] A. Fernández-Pacheco, J. M. De Teresa, J. Orna, L. Morellon, P. A. Algarabel, J. A. Pardo, M. R. Ibarra, C. Magen, and E. Snoeck, Giant planar Hall effect in epitaxial $Fe_3O_4$ thin films and its temperature dependence. Phys. Rev. B **78**, 212402 (2008).

[30] S. Deng, R. Heid, K.-P. Bohnen, C. Wang, and C. Sürgers, Minority-spin conduction in ferromagnetic $Mn_5Ge_3C_x$ and $Mn_5Si_3C_x$ films derived from anisotropic magnetoresistance and density functional theory. Phys. Rev. B **103**, 134439 (2021).

[31] D. Kriegner, K. Výborný, K. Olejník, H. Reichlová, V. Novák, X. Marti, J. Gazquez, V. Saidl, P. Němec, V. V. Volobuev, G. Springholz, V. Holý, and T. Jungwirth, Multiple-stable anisotropic magnetoresistance memory in antiferromagnetic MnTe. Nat. Commun. **7**, 11623 (2016).

[32] R. D. Gonzalez Betancourt, J. Zubáč, K. Geishendorf, P. Ritzinger, B. Růžičková, T. Kotte, J. Železný, K. Olejník, G. Springholz, B. Büchner, A. Thomas, K. Výborný, T. Jungwirth, H. Reichlová, and D. Kriegner, Anisotropic magnetoresistance in altermagnetic MnTe. npj Spintronics **2**, 45 (2024).

[33] C. Jin, P. Li, W. B. Mi, and H. L. Bai, Magnetocrystalline anisotropy-dependent six-fold

symmetric anisotropic magnetoresistance in epitaxial $Co_xFe_{3-x}O_4$ films. Europhys. Lett. **100**, 27006 (2012).

[34] L. Song, F. Zhou, J. Chen, H. Li, X. Xi, Y.-C. Lau, and W. Wang, In-plane anisotropic magnetoresistance and planar Hall effect in off-stoichiometric single crystal Mn3Ga. Appl. Phys. Lett. **125**, 102403 (2024).

[35] D. Huang, H. Nakamura, and H. Takagi, Planar Hall effect with sixfold oscillations in a Dirac antiperovskite. Phys. Rev. Res. **3**, 013268 (2021).

[36] P. K. Rout, I. Agireen, E. Maniv, M. Goldstein, and Y. Dagan, Six-fold crystalline anisotropic magnetoresistance in the (111) $LaAlO_3/SrTiO_3$ oxide interface. Phys. Rev. B **95**, 241107 (2017).

[37] R. R. Birss, *Symmetry and magnetism* (North-Holland, Amsterdam, 1964).

[38] Y. Miao, J. Sun, C. Gao, D. Xue, and X. R. Wang, Anisotropic galvanomagnetic effects in single cubic crystals: A theory and its verification. Phys. Rev. Lett. **132**, 206701 (2024).

[39] F. L. Zeng, Z. Y. Ren, Y. Li, J. Y. Zeng, M. W. Jia, J. Miao, A. Hoffmann, W. Zhang, Y. Z. Wu, and Z. Yuan, Intrinsic mechanism for anisotropic magnetoresistance and experimental confirmation in $Co_xFe_{1-x}$ single-crystal films. Phys. Rev. Lett. **125**, 097201 (2020).

[40] Y. Dai, Y. W. Zhao, L. Ma, M. Tang, X. P. Qiu, Y. Liu, Z. Yuan, and S. M. Zhou, Fourfold anisotropic magnetoresistance of $L1_0$ FePt due to relaxation time anisotropy. Phys. Rev. Lett. **128**, 247202 (2022).

[41] S. Kokado, M. Tsunoda, K. Harigaya, and A. Sakuma, Anisotropic magnetoresistance effects in Fe, Co, Ni, $Fe_4N$, and half-metallic ferromagnet: A systematic analysis. J. Phys. Soc. Jpn. **81**, 024705 (2012).

[42] J.-H. Park, H.-W. Ko, J.-M. Kim, J. Park, S.-Y. Park, Y. Jo, B.-G. Park, S. K. Kim, K.-J. Lee, and K.-J. Kim, Temperature dependence of intrinsic and extrinsic contributions to anisotropic magnetoresistance. Sci. Rep. **11**, 20884 (2021).

[43] M. Q. Dong, Z.-X. Guo, and X. R. Wang, Anisotropic magnetoresistance due to magnetization-dependent spin-orbit interactions. Phys. Rev. B **108**, L020401 (2023).

[44] K. M. Seemann, F. Freimuth, H. Zhang, S. Blügel, Y. Mokrousov, D. E. Bürgler, and C. M. Schneider, Origin of the planar Hall effect in nanocrystalline $Co_{60}Fe_{20}B_{20}$. Phys. Rev. Lett. **107**, 086603 (2011).

[45] L. Nádvorník, M. Borchert, L. Brandt, R. Schlitz, K. A. de Mare, K. Výborný, I. Mertig, G. Jakob, M. Kläui, S. T. B. Goennenwein, M. Wolf, G. Woltersdorf, and T. Kampfrath, Broadband terahertz probes of anisotropic magnetoresistance disentangle extrinsic and intrinsic contributions. Phys. Rev. X **11**, 021030 (2021).

[46] W. S. Hou, M. Q. Dong, X. Zhang, and Z.-X. Guo, Giant anisotropic magnetoresistance in magnetic monolayers $CrPX_3$ (X=S, Se, Te) due to symmetry breaking between the in-plane and out-of-plane crystallographic axes. Phys. Rev. B **110**, 214403 (2024).

[47] B. Khodadadi, A. Rai, A. Sapkota, A. Srivastava, B. Nepal, Y. Lim, D. A. Smith, C. Mewes, S. Budhathoki, A. J. Hauser, M. Gao, J.-F. Li, D. D. Viehland, Z. Jiang, J. J. Heremans, P. V. Balachandran, T. Mewes, and S. Emori, Conductivitylike Gilbert damping due to intraband scattering in epitaxial iron. Phys. Rev. Lett. **124**, 157201 (2020).

[48] See Supplemental Material at [URL] for the sample characterization, detailed analysis of experimental data, phenomenological theory of high-order AMR, and first-principles calculations, which includes Refs. [49-51].

[49] Y. Liu, A. A. Starikov, Z. Yuan, and P. J. Kelly, First-principles calculations of magnetization relaxation in pure Fe, Co, and Ni with frozen thermal lattice disorder. Phys. Rev. B **84**, 014412 (2011).

[50] C. Y. Ho, M. W. Ackerman, K. Y. Wu, T. N. Havill, R. H. Bogaard, R. A. Matula, S. G. Oh, and H. M. James, Electrical resistivity of ten selected binary alloy systems. J. Phys. Chem. Ref. Data **12**, 183 (1983).

[51] J. R. Yates, X. Wang, D. Vanderbilt, and I. Souza, Spectral and Fermi surface properties from Wannier interpolation. Phys. Rev. B **75**, 195121 (2007).

[52] Y. Miao, X. Chen, S. Yang, K. Zheng, Z. Lian, Y. Wang, P. Wang, C. Gao, D.-Z. Yang, and D.-S. Xue, Non-cosine square angular-dependent magnetoresistance of the face-centered-cubic Co thin films. J. Magn. Magn. Mater. **512**, 167013 (2020).

[53] C. Goldberg and R. E. Davis, New galvanomagnetic effect. Phys. Rev. **94**, 1121 (1954).

[54] V. A. Marsocci and T. T. Chen, Measurements of the planar Hall effect in polycrystalline and in single-crystal nickel thin films. J. Appl. Phys. **40**, 3361 (1969).

[55] T. T. Chen and V. A. Marsocci, Planar magnetoresistivity and planar Hall effect measurements in nickel single-crystal thin films. Physica **59**, 498 (1972).

[56] A. L. Ravensburg, M. Werwiński, J. Rychły-Gruszecka, J. Snarski-Adamski, A. Elsukova, P. O. Å. Persson, J. Rusz, R. Brucas, B. Hjörvarsson, P. Svedlindh, G. K. Pálsson, and V. Kapaklis, Boundary-induced phase in epitaxial iron layers. Phys. Rev. Mater. **8**, L081401 (2024).

[57] I. I. Mazin, How to define and calculate the degree of spin polarization in ferromagnets. Phys. Rev. Lett. **83**, 1427 (1999).

[58] A. A. Starikov, Y. Liu, Z. Yuan, and P. J. Kelly, Calculating the transport properties of magnetic materials from first principles including thermal and alloy disorder, noncollinearity, and spin-orbit coupling. Phys. Rev. B **97**, 214415 (2018).

[59] S. Kokado and M. Tsunoda, Twofold and fourfold symmetric anisotropic magnetoresistance effect in a model with crystal field. J. Phys. Soc. Jpn. **84**, 094710 (2015).

[60] M. Dąbrowski, T. R. F. Peixoto, M. Pazgan, A. Winkelmann, M. Cinal, T. Nakagawa, Y. Takagi, T. Yokoyama, F. Bisio, U. Bauer, F. Yildiz, M. Przybylski, and J. Kirschner, Oscillations of the orbital magnetic moment due to d-Band quantum well states. Phys. Rev. Lett. **113**, 067203 (2014).

[61] Y. Chen, H. Chen, X. Shen, W. Chen, Y. Liu, Y. Wu, and Z. Yuan, Orbital-excitation-dominated magnetization dissipation and quantum oscillation of Gilbert damping in Fe films. Phys. Rev. Lett. **134**, 136701 (2025).

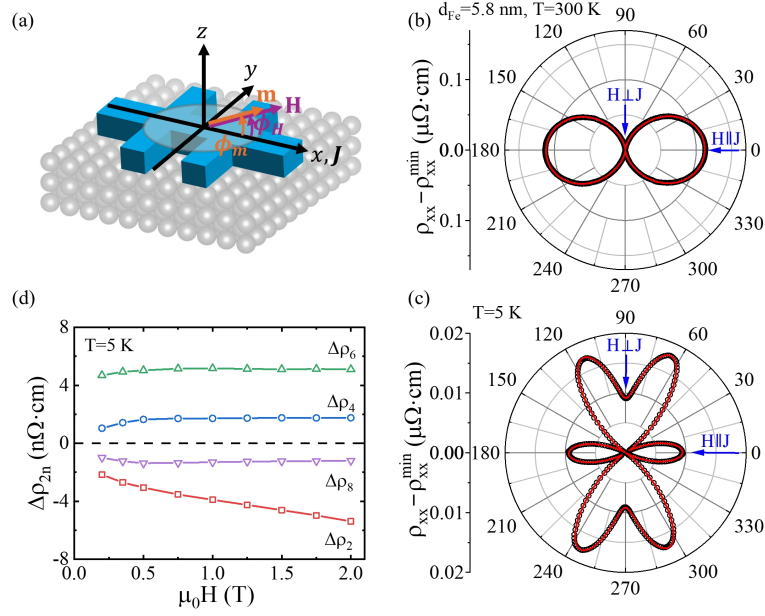

Fig. 1. (a) Schematic of in-plane angle-dependent AMR measurement using a vector magnet. The angle $\phi_H$ denotes the in-plane angle between current and applied field, while $\phi_m$ refers to the angle between current and magnetization. (b) Angular dependence of $\rho_{xx}(\phi_H)$ in a 5.8-nm Fe(001) film at 300 K. The red curve is a fit using the conventional two- and four-fold AMR expression. (c) $\rho_{xx}(\phi_H)$ measured at 5 K, exhibiting a pronounced $\cos 6\phi_H$ component. The red curve includes cosine terms up to the eighth order [Eq. (1)]. (d) Field dependence of the fitted amplitudes $\Delta\rho_4$, $\Delta\rho_6$, and $\Delta\rho_8$ saturate above 0.5 T, whereas $\Delta\rho_2$ increases linearly with field.

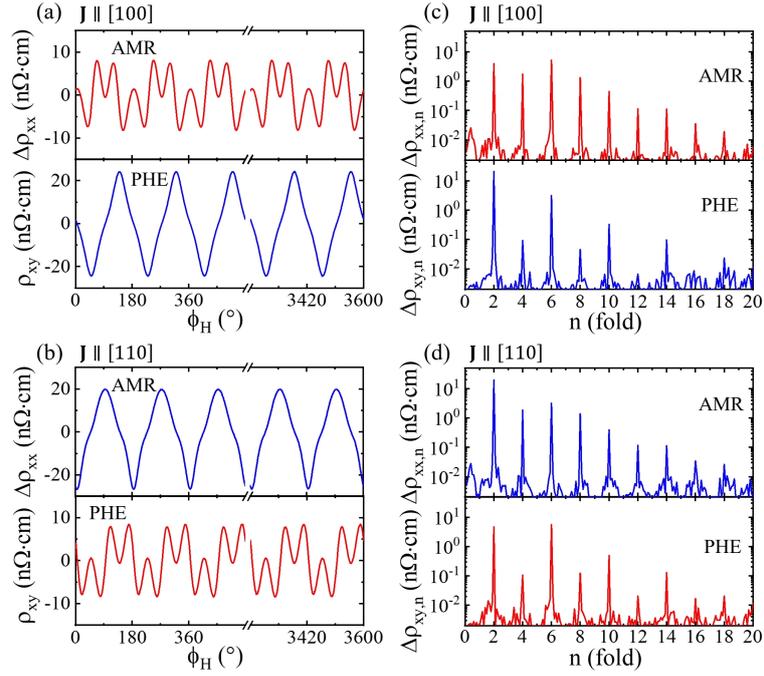

Fig. 2. (a,b) AMR and PHE for a 5.8-nm Fe(001) film with current along (a) [100] and (b) [110]. All data were taken at 5 K under a rotating field of 1 T. The correspondence between AMR in (a) and PHE in (b), and vice versa, highlights their reciprocal relation imposed by crystal symmetry. (c,d) FFT spectra of the signals in (a) and (b), showing distinct peaks up to the 18-fold component.

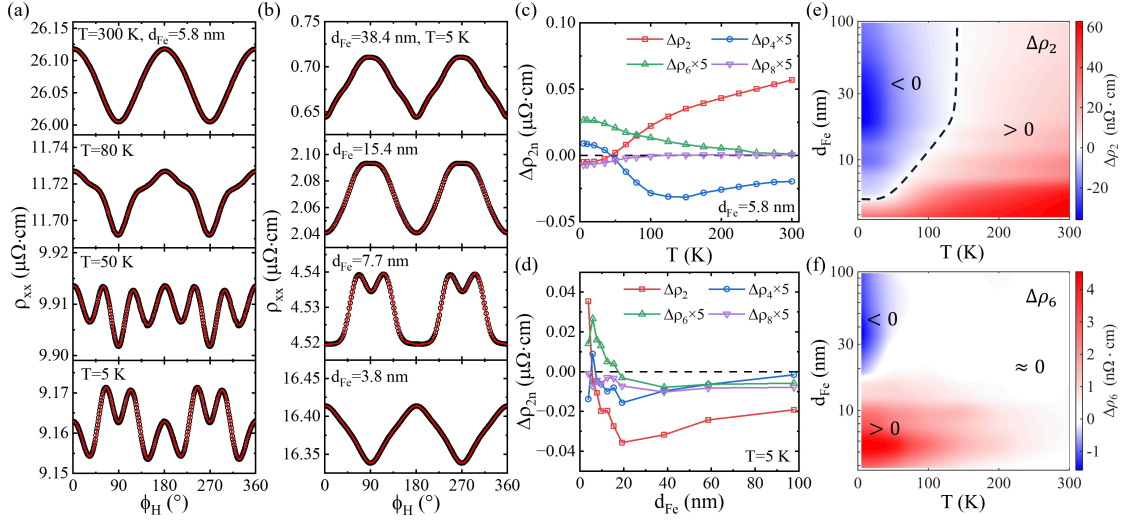

Fig. 3. (a) Temperature dependence of AMR in a 5.8-nm Fe(001) film. (b) Thickness dependence of AMR at 5 K. The rotating field strength is 1 T. In both (a) and (b), the red curves represent fits using Eq. (1). (c) Extracted coefficients $\Delta\rho_2$, $\Delta\rho_4$, $\Delta\rho_6$, and $\Delta\rho_8$ as functions of temperature for the 5.8-nm film. (d) Thickness dependence of $\Delta\rho_2$, $\Delta\rho_4$, $\Delta\rho_6$, and $\Delta\rho_8$ at 5 K. (e,f) Two-dimensional phase diagrams of $\Delta\rho_2$ and $\Delta\rho_6$ as functions of temperature and thickness, revealing the critical boundaries of sign reversal. The dashed line in (e) indicates the guideline where $\Delta\rho_2 = 0$.

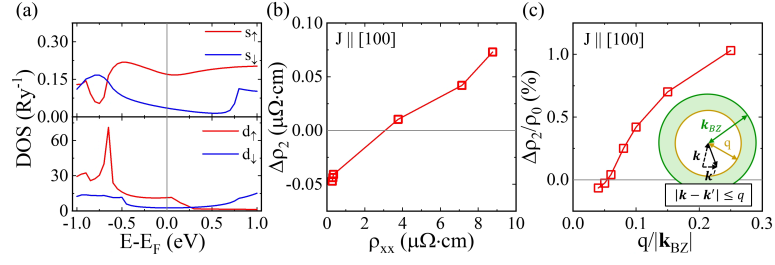

Fig. 4. (a) Spin-resolved projected DOS of *s* and *d* electrons in bcc Fe at the Fermi level. (b) $\Delta\rho_2$ obtained from full quantum-mechanical transport calculations as phonon excitation increases, showing a sign reversal with the increasing $\rho_{xx}$. (c) Minimal scattering-model calculation of $\Delta\rho_2$ using Eq. (4) as a function of cutoff wavevector q. The inset illustrates the constraint $|\mathbf{k} - \mathbf{k}'| \leq q$ in the scattering process.

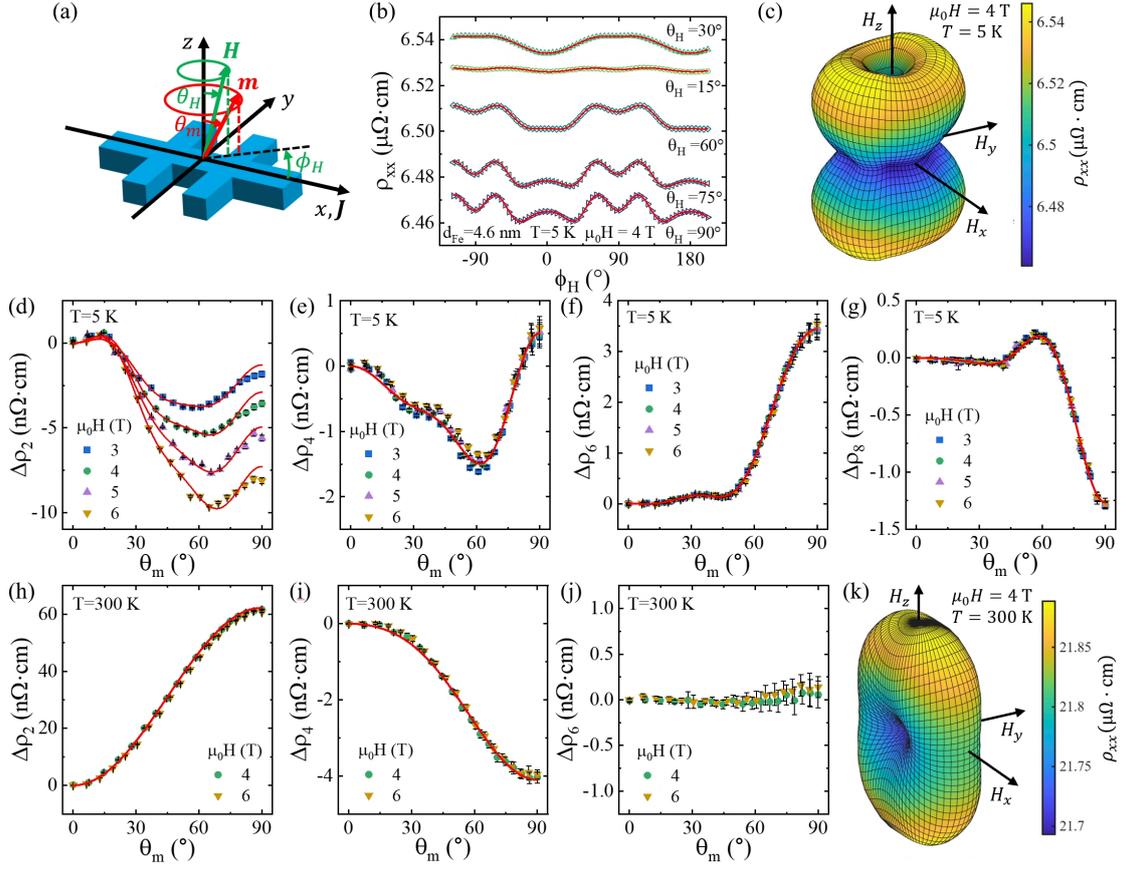

Fig. 5. (a) Schematic of full angular mapping including both azimuthal ($\phi_H$) and polar ($\theta_H$) field rotations. (b) $\rho_{xx}(\phi_H)$ at selected $\theta_H$ values in a 4.6-nm film at $T = 5$ K and $\mu_0 H = 4$ T. As $\theta_H$ decreases from 90° (in-plane) to 0° (out-of-plane), the AMR amplitude gradually diminishes. (c) Full angular map of $\rho_{xx}(\theta_H, \phi_H)$, revealing higher-order angular structures beyond the film plane. (d-g) Extracted harmonic coefficients $\Delta\rho_2$, $\Delta\rho_4$, $\Delta\rho_6$, and $\Delta\rho_8$ as functions of the magnetization polar angle $\theta_m$ at 5 K. Data in (e-g) taken at different fields collapse onto a single curve, confirming the intrinsic nature of the high-order terms. The red curves in (d-g) are fits using a series of $\sin^{2n}\theta_m$ terms up to $n = 4$. (h-j) Polar-angle dependence of $\Delta\rho_2$, $\Delta\rho_4$, and $\Delta\rho_6$ extracted from full angular-map measurements at 300 K on the same sample, showing that $\Delta\rho_6$ is nearly zero within experimental uncertainty. The red curves in (h-j) are fits using the series only up to $n = 2$. (k) Full angular map of $\rho_{xx}(\theta_H, \phi_H)$ measured at 300 K, exhibiting no high-order angular features beyond four-fold symmetry.

# Supplemental Materials for

# Observation of High-Order Anisotropic Magnetoresistance in a Cubic Ferromagnet

Haoran Chen[1, #], Yue Chen[2, 3, #], Yizi Feng[1], Ruda Guo[2, 3], Yuanfei Fan[1], Hongyue Xu[1], Tong Wu[1], Zhongxun Guo[1], Di Yue[1], Xiaofeng Jin[1], Yi Liu[6, 7], Zhe Yuan[3, †], Yizheng Wu[1,4,5, *]

[1]*Department of Physics, Applied Surface Physics State Key Laboratory, Fudan University, Shanghai 200433, China*
[2]*The Center for Advanced Quantum Studies and School of Physics and Astronomy, Beijing Normal University, Beijing 100875, China*
[3]*Interdisciplinary Center for Theoretical Physics and Information Science, Fudan University, Shanghai 200433, China*
[4]*Shanghai Research Center for Quantum Sciences, Shanghai 201315, China*
[5]*Shanghai Key Laboratory of Metasurfaces for Light Manipulation, Fudan University, Shanghai 200433, China*
[6]*Institute for Quantum Science and Technology, Shanghai University, Shanghai 200444, China*
[7]*Department of Physics, Shanghai University, Shanghai 200444, China*

## I. Characterization of crystal structure

The emergence of high-order AMR harmonics in Fe(001) films, as discussed in the main text, is fundamentally linked to the underlying crystalline symmetry. To provide a structural foundation for our transport analysis, we characterized the epitaxial quality and symmetry of Fe(001) films grown on $MgAl_2O_4$(001) substrates using x-ray diffraction (XRD).

Figure S1(a) displays the out-of-plane XRD scan around the Fe(002) reflection. In addition to the sharp film and substrate peaks, pronounced Laue oscillations are visible, indicating smooth interfaces, uniform thickness, and excellent crystallinity. The presence of such oscillations establishes that the films are coherent single crystals rather than textured polycrystalline aggregates.

We note that, due to the lattice mismatch between Fe and $MgAl_2O_4$, the epitaxial Fe(001) films exhibit a boundary-induced body-centered-tetragonal (bct) distortion, characterized by an elongation along the out-of-plane [001] direction and a corresponding compression along the in-plane [100] and [010] directions, as reported previously [1].

To probe the in-plane symmetry, we performed ϕ-scan measurements of the Fe{112} reflections. As shown in Fig. S1(b), four sharp peaks appear at 90° intervals as the sample is rotated in-plane, directly evidencing the four-fold rotational symmetry of the Fe(001) lattice inherited from the cubic bulk crystal. Although the intensities of the four reflections are not strictly identical, this asymmetry does not originate from the bct distortion itself, which preserves four-fold rotational symmetry about the film normal. Instead, the intensity difference is attributed to extrinsic factors such as a slight substrate miscut or minor sample misalignment during the XRD measurement. The absence of extra peaks rules out twinning or rotational domains.

To further quantify the bct distortion, we extracted the out-of-plane lattice constant c for films with different thicknesses. As shown in Fig. S1(c), the lattice constant c is enhanced by approximately 0.42% relative to bulk bcc Fe for a 19.2-nm-thick film and gradually relaxes toward the bulk value with increasing thickness.

These structural characterizations provide the essential basis for interpreting our magnetotransport data. Importantly, the bct distortion preserves four-fold rotational symmetry and is fully accounted for in the $C_4$ symmetry analysis adopted in this work. The confirmation of single-crystal epitaxy with well-defined four-fold symmetry ensures that the observed high-order AMR components, including the six-fold and even 18-fold harmonics reported in the main text, originate from intrinsic electronic and scattering anisotropies allowed by symmetry rather than extrinsic disorder or structural artifacts. While thermal expansion at low temperatures may slightly modify the magnitude of the bct distortion, it does not lower the crystal symmetry or break the four-fold rotational invariance relevant to our analysis.

Following structural characterization, the films were patterned into Hall bar devices (600 μm × 150 μm) using standard photolithography and Ar-ion etching. Two types of Hall bars were fabricated, with the current directed along Fe[100] and Fe[110]. The longitudinal and transverse voltages were measured using a standard lock-in technique, with an AC excitation frequency of 137.31 Hz. Depending on film thickness, the electric current amplitude ranged from 1 to 5 mA, small enough to avoid Joule heating during measurement.

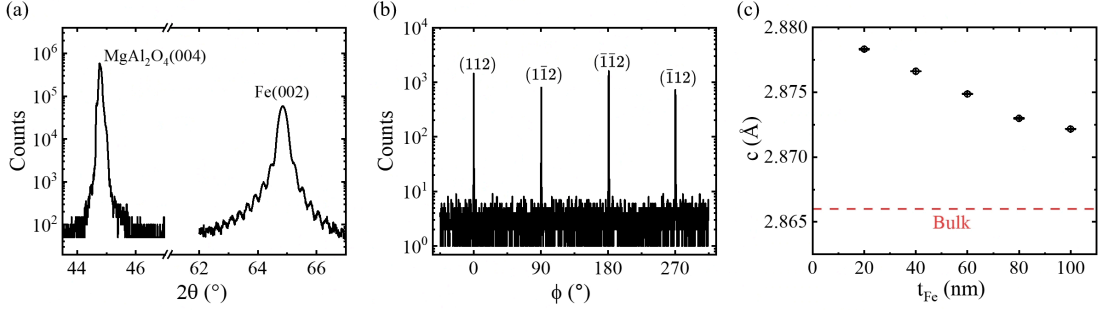

Fig. S1. (a) Out-of-plane XRD scan of a 38.4 nm Fe(001) film grown on MgAl$_2$O$_4$(001). The sharp Fe(002) peak and pronounced Laue oscillations demonstrate excellent crystallinity, smooth interfaces, and uniform thickness. (b) In-plane XRD ϕ-scan of the Fe{112} reflections. Four peaks separated by 90° directly evidence the four-fold rotational symmetry of the Fe(001) lattice, consistent with the symmetry assumed in the main-text analysis. (c) Thickness dependence of the out-of-plane lattice constant c, showing a boundary-induced bct distortion that gradually relaxes toward the bulk bcc value with increasing film thickness.

## II. Fitting order analysis of angular-dependent AMR

To verify the robustness of the extracted angular harmonics, we tested the fitting of the angular-dependent AMR data for the 5.8-nm Fe(001) film at 5 K using cosine series truncated at different maximum orders. Figure S2(a) shows fits to Eq. (1) using maximum orders ranging from 4 to 8. At low fitting orders, systematic deviations from the experimental data are visible, particularly around the extrema, reflecting the missing higher-order contributions. The corresponding residuals, defined as the difference between measured and fitted values, are plotted in Fig. S2(b). As the fitting order increases, the residual magnitude decreases, and for truncations at the 8th order or higher, the residuals become negligible and structureless, indicating that all physically relevant harmonics have been included.

To further quantify the fitting quality, we calculated the coefficient of determination ($R^2$) as a function of fitting order [Fig. S2(c)]. The $R^2$ value increases monotonically toward 1.0 and saturates beyond the 8th order, confirming that additional terms beyond this point do not improve the fit. Importantly, owing to the orthogonality of the cosine basis functions, the extracted amplitudes of the lower-order harmonics remain essentially unchanged regardless of the truncation order. This ensures that the determination of the low-order terms is robust and unaffected by the inclusion or exclusion of higher-order components.

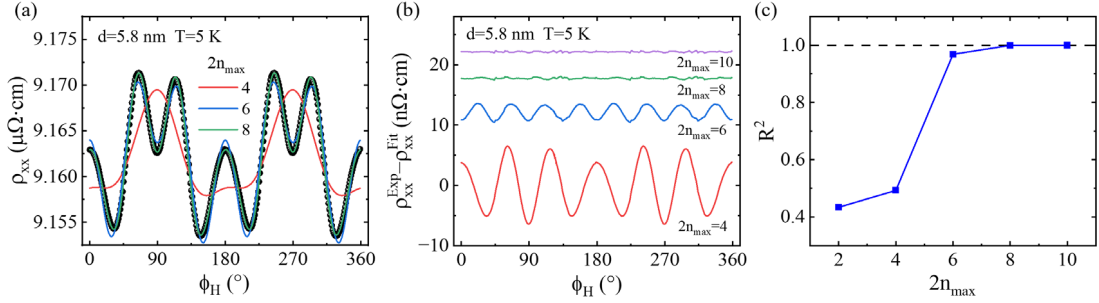

Fig. S2. (a) Fits of the angular-dependent AMR using Eq. (1) truncated at maximum orders of 4, 6, and 8. (b) Residuals (measured minus fitted values) for different fitting orders, vertically offset for clarity. (c) Coefficient of determination ($R^2$) as a function of fitting order, showing saturation above the 8th order.

### III. Field-to-magnetization angle conversion

In field-rotation measurements, the experimental angles correspond to the applied magnetic field direction $(\theta_H, \phi_H)$. However, due to magnetic anisotropy, the magnetization $(\theta_m, \phi_m)$ does not perfectly align with the field, particularly during out-of-plane rotation. For quantitative analysis of AMR, it is therefore necessary to convert field angles into the actual magnetization angles [2].

The magnetic anisotropy of Fe(001) films can be modeled by two contributions [Fig. S3(a,e)]: a strong out-of-plane uniaxial anisotropy (dominated by demagnetization) and a weaker cubic anisotropy from the crystal field. Including the Zeeman term, the total magnetic energy in Fe(001) film is given by

$$\begin{aligned}
\mathcal{E} &= \mathcal{E}_z + \mathcal{E}_c + \mathcal{E}_u, \\
\mathcal{E}_z &= -\mu_0 M_s (H_x m_x + H_y m_y + H_z m_z), \\
\mathcal{E}_c &= \frac{1}{2} \mu_0 M_s H_c (m_x^2 m_y^2 + m_x^2 m_z^2 + m_y^2 m_z^2), \\
\mathcal{E}_u &= -\frac{1}{2} \mu_0 M_s H_{u1} m_z^2.
\end{aligned} \qquad (S1)$$

where $\mu_0 H_c \approx 0.05$ T is the cubic anisotropy field, and $\mu_0 H_{u1} \approx 2.1$ T is the out-of-plane uniaxial anisotropy field. The magnetization vector is parameterized as $m = (m_x, m_y, m_z) = (\sin\theta_m \cos\phi_m, \sin\theta_m \sin\phi_m, \cos\theta_m)$.

The mapping from $(\theta_H, \phi_H)$ to $(\theta_m, \phi_m)$ is obtained by numerical minimization of Eq. (S1). Two representative cases are summarized below:

   **(i) Out-of-plane rotation.** — Figure S3(b) illustrates the measurement geometry. The mapping $\theta_H$ to $\theta_m$ at $\mu_0 H = 4$ T is given in Fig. S3(c), with the deviation $\theta_m - \theta_H$ plotted in Fig. S3(d). The misalignment can exceed 15°, highlighting the importance of converting $\theta_H$ to $\theta_m$. Because $H_{u1} \gg H_c$, the cubic anisotropy is negligible in this configuration. These corrections are applied to the out-of-plane AMR analysis in Fig. 5 of the main text.

   **(ii) In-plane rotation.** — Figure S3(f) shows the geometry. The mapping $\phi_H$ to $\phi_m$ at $\mu_0 H = 1$ T is shown in Fig. S3(g), with the deviation $\phi_H - \phi_m$ plotted in Fig. S3(h). The difference is less than 0.75°, justifying the approximation $\phi_H \approx \phi_m$.

   In summary, conversion from $\theta_H$ to $\theta_m$ is essential when the field is tilted out of plane, whereas conversion from $\phi_H$ to $\phi_m$ is negligible at $\mu_0 H = 1$ T. The latter correction only becomes relevant when identifying subtle high-order harmonics in the presence of a dominant two-fold AMR contribution, as discussed in the next section.

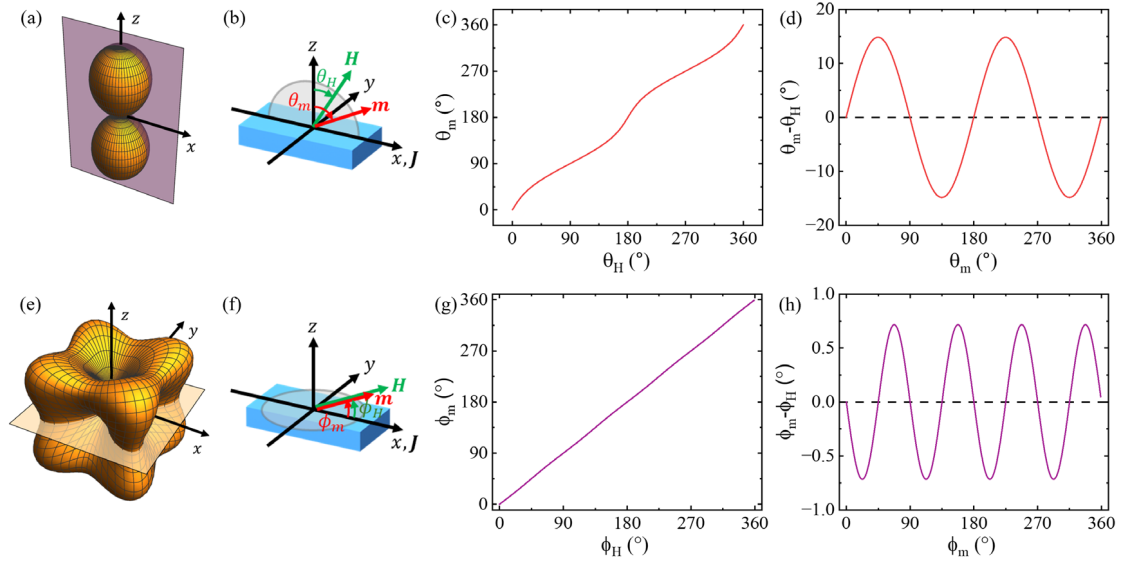

Fig. S3. (a) Illustration of out-of-plane uniaxial anisotropy. (b) Geometry for out-of-plane rotation. (c) Mapping between $\theta_H$ and $\theta_m$ at $\mu_0 H = 4$ T. (d) Angular deviation $\theta_m - \theta_H$ during out-of-plane rotation. (e) Illustration of cubic magnetic anisotropy. (f) Schematic of the in-plane rotation geometry. (g) Mapping between $\phi_H$ and $\phi_m$ at $\mu_0 H = 1$ T. (h) Angular deviation $\phi_m - \phi_H$ during in-plane rotation, showing a misalignment below 0.75°.

## IV. Effect of field-magnetization misalignment on apparent AMR harmonics

In the previous section, we demonstrated how the applied field angles $(\theta_H, \phi_H)$ are converted into the actual magnetization angles $(\theta_m, \phi_m)$. For in-plane rotations at $\mu_0 H = 1$ T, the misalignment between the field and magnetization is below 0.75°, so the conversion is often negligible. Nevertheless, even a small deviation can generate artificial higher-order harmonics in the $\phi_H$-dependent AMR, even if the intrinsic AMR depends purely on $\phi_m$ through a two-fold term.

To illustrate this effect, we assume a pure two-fold intrinsic AMR $\rho_{xx}(\phi_m) = \cos 2\phi_m$, and compute the apparent $\rho_{xx}(\phi_H)$ using the relation between $\phi_m$ and $\phi_H$ obtained in the previous section. These simulated curves represent experimental observations under finite field-magnetization misalignment. Figure S4(a) shows the calculated $\rho_{xx}(\phi_H)$ for different field-to-anisotropy ratios $H_{ext}/H_c$ ranging from 2 to 80. At small ratios, the curves deviate visibly from the ideal cosine form, particularly near extrema, reflecting the emergence of spurious high-order contributions. The corresponding Fourier coefficients $\Delta\rho_2$, $\Delta\rho_4$, $\Delta\rho_6$, and $\Delta\rho_8$ extracted from fits to Eq. (1) are summarized in Fig. S4(b). As expected, $\Delta\rho_4$ and $\Delta\rho_8$ vanish for all fields, whereas a false six-fold term $\Delta\rho_6$ appears at low fields and gradually decays to zero as the field strength increases. The $\Delta\rho_2$ term also deviates slightly from unity at low fields but converges to its intrinsic value as $H_{ext}/H_c$ increases.

In our experiment, the applied field is 1 T and the in-plane cubic anisotropy field is ~0.05 T, yielding $H_{ext}/H_c \approx 20$. Under these conditions, an intrinsic AMR of $\rho_{xx}(\phi_m) = \cos 2\phi_m$ produces an apparent $\phi_H$-dependent form $\rho_{xx}(\phi_H) = \rho_0 + 1.012\Delta\rho_2 \cos 2\phi_H - 0.013\Delta\rho_2 \cos 6\phi_H$. Thus, a 1 T rotating field can induce a weak artificial six-fold term (~1.3% of the two-fold amplitude) solely due to the small misalignment between field and magnetization.

Using the calculated $\theta_m - \theta_H$ relation, we further converted the measured $\rho_{xx}(\phi_H)$ data into $\rho_{xx}(\phi_m)$ and refitted them using $\rho_{xx}(\phi_m) = \rho_0 + \sum_{n=1}^{4} \Delta\rho_{2n} \cos(2n\phi_m)$. Figures S4(c) and S4(d) compare the temperature-dependent $\Delta\rho_2$ and $\Delta\rho_6$ obtained with and without the angular correction. The correction has little effect on $\Delta\rho_2$ across all temperatures (within 1.2%), but it significantly alters $\Delta\rho_6$ near room temperature due to the large $\Delta\rho_2$ component. When uncorrected, a small negative $\Delta\rho_6$ appears at high temperatures, whereas the corrected data show $\Delta\rho_6 \to 0$. At low temperatures, however, the correction is negligible since $\Delta\rho_6$ is comparable in magnitude to $\Delta\rho_2$.

Taken together, these analyses demonstrate that while field-magnetization misalignment can generate a small artificial six-fold AMR at high temperatures, the pronounced high-order harmonics observed at 5 K in the main text are intrinsic and robust against such artifacts.

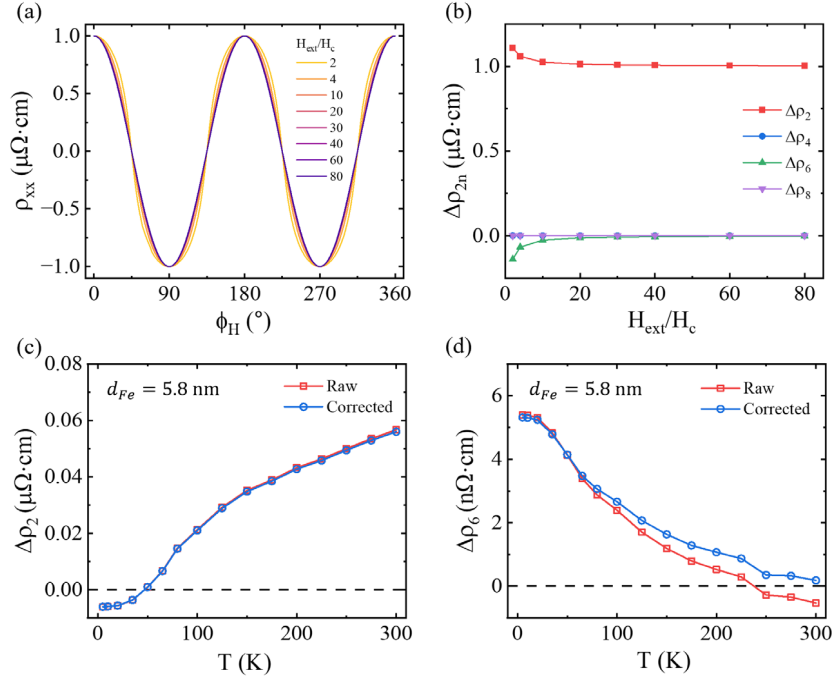

Fig. S4. (a) Simulated $\rho_{xx}(\phi_H)$ curves assuming intrinsic two-fold AMR $\rho_{xx}(\phi_m) = \cos 2\phi_m$, with different field-to-anisotropy ratios $H_{ext}/H_c$. (b) Extracted harmonic coefficients as functions of $H_{ext}/H_c$. Artificial six-fold contributions appear at low fields but vanish as field strength increases. (c,d) Temperature dependence of experimental $\Delta\rho_2$ and $\Delta\rho_6$ obtained with (blue) and without (red) the $\phi_H$ to $\phi_m$ conversion. The correction has little effect on $\Delta\rho_2$ but removes the spurious negative $\Delta\rho_6$ at 300 K.

**V. Subtracting OMR effect and identifying the negative intrinsic AMR at low temperature.**

As discussed in the main text, early studies attributed the negative AMR observed in Fe films at low temperatures to ordinary magnetoresistance (OMR) arising from the Lorentz force [3,4]. To clarify its contribution, we performed field-sweep measurements at 5 K. Figure S5(a) shows the longitudinal resistivity $\rho_{xx}$ of a 97.7-nm Fe film as a function of the applied magnetic field, with the field oriented along either the x-axis (blue) or y-axis (green). The difference in zero-field resistivity for fields applied along [100] and [010] arises from the intrinsic AMR associated with magnetization aligned along the two orthogonal easy axes of Fe(001). The different slopes for the two orientations indicate the presence of an additional OMR contribution, superimposed on the intrinsic AMR measured under in-plane rotation at $\mu_0 H = 1$ T.

In the OMR mechanism, the Lorentz force is proportional to the magnetic induction $B = \mu_o(H + M)$. To isolate this effect, Fig. S5(b) replots $\rho_{xx}$ as a function of B and fits it linearly (red line). The intercept of the fit represents the intrinsic AMR component at B = 0 T, where the field-induced OMR contribution is eliminated. Even after subtracting OMR, $\rho_{xx}(H \parallel y)$ remains larger than $\rho_{xx}(H \parallel x)$, revealing an intrinsic origin of the negative $\Delta\rho_2$ beyond OMR. The extracted OMR coefficients for $H \parallel x$ and $H \parallel y$ are summarized in Fig. S5(c), showing pronounced anisotropy, and OMR is significantly stronger when the field is applied along the y-axis.

Figure S5(d) presents the thickness dependence of $\Delta\rho_2$ at 5 K. The black curve shows the raw data obtained at B = 3.1 T, while the red curve corresponds to the corrected values at B = 0 T after subtracting OMR using the procedure above. Although subtraction reduces the magnitude of the negative AMR in thick films, it does not reverse its sign, demonstrating that OMR enhances—but does not fully account for—the observed negative $\Delta\rho_2$ contribution. Thus, unlike earlier interpretations that attributed the negative AMR entirely to OMR, our results indicate that the intrinsic AMR itself becomes negative in thick Fe films at low temperatures.

Finally, we note that the OMR contribution is negligible for higher-order harmonics: as shown in Fig. 1(d) of the main text, $\Delta\rho_4$, $\Delta\rho_6$, and $\Delta\rho_8$ all saturate above 0.5 T, confirming that Lorentz-force-driven OMR does not affect the observed high-order AMR.

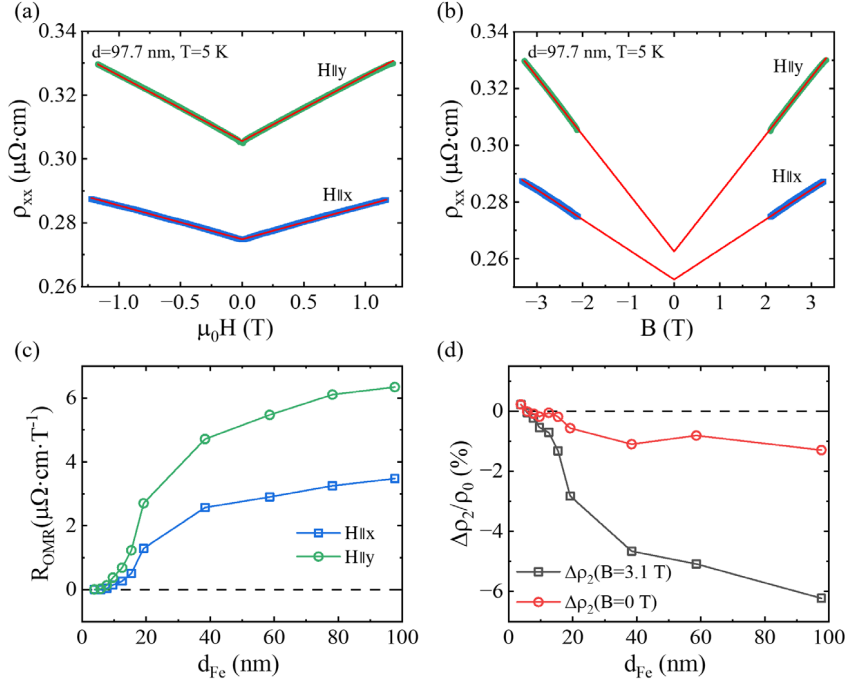

Fig. S5. (a) Longitudinal resistivity $\rho_{xx}$ versus applied field $\mu_0 H$ along the x-axis (blue) and y-axis (green) at 5 K. (b) $\rho_{xx}$ replotted as a function of magnetic induction $B = \mu_o(H + M)$ with a linear fit (red). The intercept yields the intrinsic AMR after removing the OMR contribution. (c) Extracted OMR coefficients for $H \parallel x$ (blue) and $H \parallel y$ (green), showing strong anisotropy with larger OMR for fields along y. (d) Thickness dependence of $\Delta\rho_2$ at 5 K. Raw data (black) and OMR-corrected values (red) are compared. Subtraction reduces but does not eliminate the negative AMR, indicating an intrinsic contribution.

## VI. Multi-rotation FFT analysis of AMR

In the main text, we reported multi-rotation measurements of $\rho_{xx}$ and $\rho_{xy}$ for a 5.8-nm-thick Fe(001) film with $J \parallel [100]$ and $J \parallel [110]$, followed by Fourier analysis of the angular harmonics. The extracted amplitudes of each harmonic peak are summarized in Table S1.

Equation (3) in the main text establishes the reciprocal relations between AMR and PHE for higher-order harmonics: $\rho_{xx}^{[100]}$ and $\rho_{xy}^{[110]}$ share the same coefficients $\Delta\rho_{4n-2}$, while $\rho_{xx}^{[110]}$ and $\rho_{xy}^{[100]}$ share the corresponding $\Delta\rho_{4n-2}^*$. In addition, $\rho_{xx}^{[100]}$ and $\rho_{xx}^{[110]}$ contain identical $\Delta\rho_{4n}$ terms.

As summarized in Table S1, the extracted FFT amplitudes for the 5.8-nm Fe(001) film confirm these reciprocal relations up to the 12th-order harmonics, with the corresponding pairs highlighted in matching colors. At higher orders, the relations become less apparent due to the signal approaching the experimental noise floor.

Table S1. Extracted amplitudes of FFT peaks for AMR and PHE harmonics in the 5.8 nm Fe(001) film with $J \parallel [100]$ and $J \parallel [110]$. Reciprocal pairs are highlighted in matching colors. Units: $n\Omega \cdot cm$.

| Order | $\Delta\rho_{xx}^{J\parallel[100]}$ | $\Delta\rho_{xy}^{J\parallel[100]}$ | $\Delta\rho_{xx}^{J\parallel[110]}$ | $\Delta\rho_{xy}^{J\parallel[110]}$ |
|---|---|---|---|---|
| 2 | 3.918 | 20.643 | 19.448 | 4.668 |
| 4 | 1.724 | 0.092 | 1.841 | 0.105 |
| 6 | 5.169 | 3.132 | 3.162 | 5.566 |
| 8 | 1.308 | 0.045 | 1.384 | 0.124 |
| 10 | 0.438 | 0.327 | 0.392 | 0.491 |
| 12 | 0.113 | 0.007 | 0.117 | 0.020 |
| 14 | 0.110 | 0.097 | 0.112 | 0.127 |
| 16 | 0.035 | 0.003 | 0.034 | 0.016 |
| 18 | 0.019 | 0.023 | 0.026 | 0.020 |

To further confirm the presence of high-order angular harmonics in thicker Fe(001) films, we performed extended in-plane rotation measurements on a 78.6-nm sample at 5 K. The magnetic field was continuously rotated through ten full cycles (a total of 3600°), providing sufficient frequency resolution for high-order Fourier analysis. Figures S6(a) and S6(b) display the raw AMR and PHE curves for $J \parallel [100]$ and $J \parallel [110]$, respectively. Although the six-fold harmonic is difficult to identify directly from the raw data, the corresponding FFT spectra in Figs. S6(c) and S6(d) reveal distinct peaks up to the 18-fold component, well above the noise floor, thereby confirming the robustness of the high-order harmonics in thicker films.

The harmonic amplitudes extracted from the 78.6-nm sample are summarized in Table S2. Reciprocal relations between AMR and PHE remain evident for higher-order terms, consistent with the symmetry analysis presented in the main text. These results demonstrate that high-order AMR harmonics are not confined to ultrathin films but persist in thick epitaxial Fe layers, establishing their intrinsic and symmetry-allowed origin.

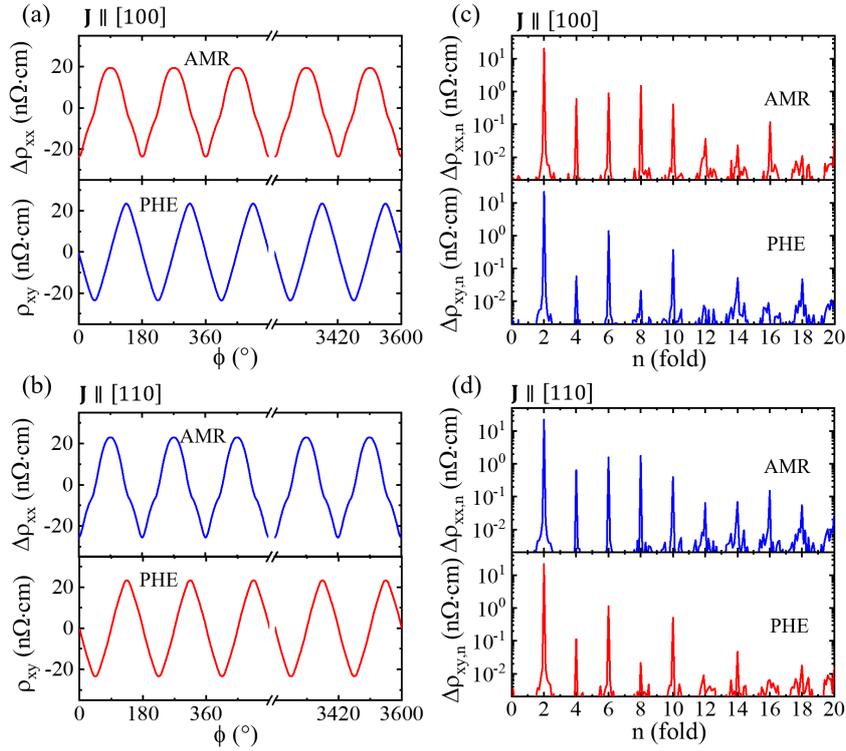

Fig. S6. (a,b) Angular dependence of AMR (upper) and PHE (lower) for Hall bars with $J \parallel [100]$ (a) and $J \parallel [110]$ in a 78.6-nm-thick Fe(001) film at 5 K. (c,d) Corresponding FFT spectra of the data in (a,b), showing distinct peaks up to the 18-fold harmonic clearly above the noise floor.

Table S2. Extracted amplitudes of FFT peaks for AMR and PHE harmonics in the 78.6 nm Fe(001) film with $J \parallel [100]$ and $J \parallel [110]$. Reciprocal pairs are highlighted in matching colors. Units: $n\Omega \cdot cm$.

| Order | $\Delta\rho_{xx}^{J\parallel[100]}$ | $\Delta\rho_{xy}^{J\parallel[100]}$ | $\Delta\rho_{xx}^{J\parallel[110]}$ | $\Delta\rho_{xy}^{J\parallel[110]}$ |
|---|---|---|---|---|
| 2 | 20.253 | 21.676 | 22.169 | 21.742 |
| 4 | 0.594 | 0.057 | 0.632 | 0.113 |
| 6 | 0.892 | 1.386 | 1.569 | 1.135 |
| 8 | 1.481 | 0.021 | 1.733 | 0.021 |
| 10 | 0.408 | 0.367 | 0.396 | 0.502 |
| 12 | 0.037 | 0.006 | 0.064 | 0.002 |
| 14 | 0.023 | 0.050 | 0.070 | 0.047 |
| 16 | 0.117 | 0.002 | 0.150 | 0.006 |
| 18 | 0.011 | 0.047 | 0.055 | 0.018 |

**VII. Measured MR ratio as a function of temperature and thickness**

In the main text, we presented the temperature and thickness dependence of the AMR components $\Delta\rho_{2n}$. Here, we further show the corresponding dependence of the normalized AMR ratios, $\Delta\rho_{2n}/\rho_0$. This representation highlights the relative contribution of each harmonic component to the total resistivity.

Figure S7(a) shows the temperature dependence of the AMR ratios for the 5.8-nm Fe(001) film, corresponding to the data in Fig. 3(c) of the main text. As temperature increases from 5 K to 300 K, the two-fold term undergoes a clear sign reversal near 50 K, while the six-fold and eight-fold components gradually decrease toward zero. The four-fold term again exhibits nonmonotonic behavior with temperature, although its magnitude remains smaller than those of the two-fold and six-fold terms.

Figure S7(b) presents the thickness dependence of $\Delta\rho_{2n}/\rho_0$ at 5 K, corresponding to Fig. 3(d) of the main text. The two-fold ratio changes sign near 5 nm, whereas the six-fold ratio reverses sign around 20 nm.

Figures S7(c) and S7(d) summarize the complete temperature and thickness dependences of $\Delta\rho_2/\rho_0$ and $\Delta\rho_6/\rho_0$, respectively, directly corresponding to the phase diagrams in Figs. 3(e) and 3(f) of the main text. The overall qualitative trends remain unchanged, confirming that the observed sign reversals and distinct film- and bulk-related contributions are robust against normalization.

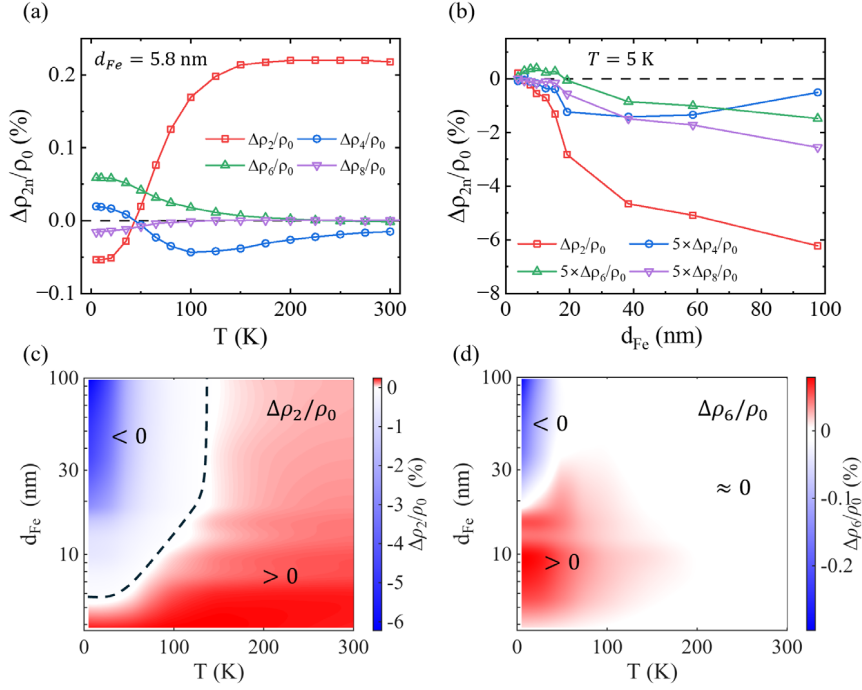

Fig. S7. (a) Temperature dependence of normalized AMR components $\Delta\rho_n/\rho_0$ in the 5.8-nm Fe(001) film. (b) Thickness dependence of normalized harmonics at 5 K. (c-d) Two-dimensional maps of $\Delta\rho_2/\rho_0$ (c) and $\Delta\rho_6/\rho_0$ (d) as functions of temperature and thickness. The dashed line in (c) (or (d)) indicates the guide line where $\Delta\rho_2/\rho_0 = 0$ (or $\Delta\rho_6/\rho_0 = 0$).

## VIII. Phenomenological theory with $C_{4v}$ symmetry

In the main text, we presented a phenomenological framework which demonstrates that two-fold, four-fold, and even higher-order AMR harmonics are symmetry-allowed in cubic (001) films. For clarity of presentation, only the final results were shown there (Eq. (3) in the main text). Here, we provide the detailed derivation of these expressions.

In conventional phenomenological theory [5-7], the resistivity tensor $\rho_{ij}$ is expanded as a Maclaurin series in the Cartesian components of the magnetization, which is then converted into angular form. While adequate for the lowest-order terms, this procedure quickly becomes cumbersome for higher-order harmonics. To treat the problem more directly, we instead write the in-plane resistivity tensor as a function of the magnetization angle $\phi_{mc}$ relative to the [100] axis:

$$\hat{\rho}(\phi_{mc}) = \begin{bmatrix} \rho_{11}(\phi_{mc}) & \rho_{12}(\phi_{mc}) \\ \rho_{21}(\phi_{mc}) & \rho_{22}(\phi_{mc}) \end{bmatrix}. \tag{S1}$$

The tensor must remain invariant under the symmetry operations of the $C_{4v}$ point group. Fourfold rotational symmetry requires $(\hat{c}_{4z}^1)^{-1} \hat{\rho}\left(\phi_{mc} + \frac{\pi}{2}\right) \hat{c}_{4z}^1 = \hat{\rho}(\phi_{mc})$ with $\hat{c}_{4z}^1 = \begin{bmatrix} 0 & -1 \\ 1 & 0 \end{bmatrix}$, yielding the constraints

$$\rho_{22}(\phi_{mc}) = \rho_{11}\left(\phi_{mc} + \frac{\pi}{2}\right), \tag{S2}$$

$$\rho_{21}(\phi_{mc}) = -\rho_{12}\left(\phi_{mc} + \frac{\pi}{2}\right). \tag{S3}$$

Mirror reflection further requires $(\sigma_v^{0°})^{-1}\hat{\rho}(-\phi_{mc})\sigma_v^{0°} = \hat{\rho}(\phi_{mc})$ with $\sigma_v^{0°} = \begin{bmatrix} -1 & 0 \\ 0 & 1 \end{bmatrix}$, yielding the constraints

$$\rho_{11}(\phi_{mc}) = \rho_{11}(-\phi_{mc}) \tag{S4}$$

$$\rho_{12}(\phi_{mc}) = -\rho_{12}(-\phi_{mc}) \tag{S5}$$

Finally, Onsager reciprocity $\rho_{ij}(m) = \rho_{ji}(-m)$ imposes

$$\rho_{ij}(\phi_{mc}) = \rho_{ji}(\phi_{mc} + \pi). \tag{S6}$$

Because $\rho_{11}(\phi_{mc})$ and $\rho_{12}(\phi_{mc})$ are periodic functions, they can be expanded in Fourier series. The symmetry constraints lead to the following restrictions on the angle-dependent terms in the Fourier series:

1. From mirror reflection [Eq. (S4)], $\rho_{11}(\phi_{mc})$ contains only cosine terms.
2. From mirror reflection [Eq. (S5)], $\rho_{12}(\phi_{mc})$ contains only sine terms.
3. From Onsager reciprocity [Eq. (S6)], $\rho_{11}(\phi_{mc})$ retains only even-order harmonics.
4. Considering rotation symmetry [Eq. (S3)] and Onsager reciprocity [Eq. (S6)], $\rho_{12}(\phi_{mc})$ must satisfy $\rho_{12}(\phi_{mc}) = -\rho_{12}(\phi_{mc} + \pi/2)$ and retain only $(4n - 2)$-order harmonics.

Thus, we find

$$\rho_{11}(\phi_{mc}) = \rho_0 + \sum_{n=1}^{\infty} \Delta\rho_{2n} \cos 2n\phi_{mc}, \tag{S7}$$

$$\rho_{12}(\phi_{mc}) = \sum_{n=1}^{\infty} \Delta\rho^*_{4n-2} \sin(4n-2)\phi_{mc}. \tag{S8}$$

***Case of current along [100] and [110]***. —With current along [100], the measured longitudinal and transverse resistivities are simply $\rho_{xx}^{[100]}(\phi_m) = \rho_{11}(\phi_m)$ and $\rho_{xy}^{[100]}(\phi_m) = \rho_{12}(\phi_m)$. With current along [110], the current direction is $\hat{l} = (\sqrt{2}/2, \sqrt{2}/2)$ and the transverse direction for Hall measurement is $\hat{t} = (-\sqrt{2}/2, \sqrt{2}/2)$. In this case, the magnetization angle is shifted as $\phi_{mc} = \phi_m + \pi/4$. Projecting the tensor components gives $\rho_{xx}^{[110]}(\phi_m) = \hat{l}_i \rho_{ij}(\phi_{mc}) \hat{l}_j = \frac{1}{2}[\rho_{11}(\phi_m) + \rho_{12}(\phi_m) + \rho_{21}(\phi_m) + \rho_{22}(\phi_m)]$ and $\rho_{xy}^{[110]}(\phi_m) = \hat{t}_i \rho_{ij}(\phi_{mc}) \hat{l}_j = \frac{1}{2}[-\rho_{11}(\phi_m) - \rho_{12}(\phi_m) + \rho_{21}(\phi_m) + \rho_{22}(\phi_m)]$. Using Eqs. (S2-S3, S7-S8), we obtain the final forms:

$$\begin{aligned}
\rho_{xx}^{[100]}(\phi_m) &= \rho_0 + \sum_{n=1}^{\infty} \Delta\rho_{4n-2} \cos(4n-2)\phi_m + \sum_{n=1}^{\infty} \Delta\rho_{4n} \cos 4n\phi_m, \\
\rho_{xy}^{[100]}(\phi_m) &= \sum_{n=1}^{\infty} \Delta\rho^*_{4n-2} \sin[(4n-2)\phi_m], \\
\rho_{xx}^{[110]}(\phi_m) &= \rho_0 + \sum_{n=1}^{\infty} (-1)^{n+1}\Delta\rho^*_{4n-2} \cos(4n-2)\phi_m + \sum_{n=1}^{\infty} (-1)^n \Delta\rho_{4n} \cos 4n\phi_m, \\
\rho_{xy}^{[110]}(\phi_m) &= \sum_{n=1}^{\infty} (-1)^{n+1}\Delta\rho_{4n-2} \sin(4n-2)\phi_m.
\end{aligned} \tag{S9}$$

Equation (S9) reproduces Eq. (3) of the main text. The detailed derivation shown here makes explicit how crystal symmetry constrains the functional form of AMR and PHE, and, importantly, why higher-order harmonics such as $\cos 6\phi_m$ and $\cos 10\phi_m$ are fully allowed by symmetry in cubic (001) films. These results provide the theoretical foundation for the experimental observations of high-order AMR reported in the main text.

***General case of arbitrary current orientation***. — For an arbitrary current direction $\phi_J$ relative to [100], the current direction is $\hat{l}(\phi_J) = (\cos\phi_J, \sin\phi_J)$ and the transverse direction for Hall measurement is $\hat{t}(\phi_J) = (-\sin\phi_J, \cos\phi_J)$. In this case, the magnetization angle is shifted as $\phi_{mc} = \phi_m + \phi_J$. Projecting the tensor components gives $\rho_{xx}(\phi_m, \phi_J) = \hat{l}_i(\phi_J)\rho_{ij}(\phi_{mc})\hat{l}_j(\phi_J)$ and $\rho_{xy}(\phi_m, \phi_J) = \hat{t}_i(\phi_J)\rho_{ij}(\phi_{mc})\hat{l}_j(\phi_J)$. Using Eqs. (S1-S3, S7-S8), we obtain:

$$\rho_{xx}(\phi_m, \phi_J) = \rho_0 + \sum_{n=1}^{\infty} \Delta\rho_{4n} \cos 4n(\phi_m + \phi_J)$$

$$+ \sum_{n=1}^{\infty} \Delta\rho_{4n-2} \cos 2\phi_J \cos[(4n-2)(\phi_m + \phi_J)] \quad (S10)$$

$$+ \sum_{n=1}^{\infty} \Delta\rho^*_{4n-2} \sin 2\phi_J \sin[(4n-2)(\phi_m + \phi_J)],$$

$$\rho_{xy}(\phi_m, \phi_J) = -\sum_{n=1}^{\infty} \Delta\rho_{4n-2} \sin 2\phi_J \cos[(4n-2)(\phi_m + \phi_J)]$$

$$+ \sum_{n=1}^{\infty} \Delta\rho^*_{4n-2} \cos 2\phi_J \sin[(4n-2)(\phi_m + \phi_J)]. \quad (S11)$$

Equations (S10) and (S11) generalize the previous model in Ref. [8], which considered only up to the four-fold symmetry terms. The framework presented here establishes the theoretical foundation for interpreting the high-order AMR harmonics observed in epitaxial Fe(001) films. It also provides a general formulation applicable to arbitrary current orientations, enabling a unified description of AMR and PHE under $C_{4v}$ symmetry.

## IX. The fully relativistic quantum-mechanical transport calculations

Within the Landauer-Büttiker scattering formalism, we constructed a system comprising a bcc Fe bulk sandwiched between two semi-infinite Au leads. The Kohn-Sham atomic sphere potentials for Fe and Au were calculated self-consistently using the tight-binding linear muffin-tin orbital method [9]. The frozen thermal lattice disorder was introduced into a 5×5 lateral supercell by displacing Fe atoms randomly from their equivalent sites according to a Gaussian distribution [10]. Periodic boundary conditions were applied along the lateral directions of the supercell. The propagating

Bloch states that are well-defined in the Au leads were incident towards the scattering region, and they were partially reflected by the disordered bcc Fe while the remaining were transmitted through the scattering region and entered the other Au lead. Employing the wave-function matching technique, we directly calculated the scattering matrix **S**, which relates the incoming and outgoing states via the reflection and transmission matrices, **r** and **t**. The total resistance $R$ of the system was then evaluated as

$$R = \left[\frac{e^2}{h}Tr(\bm{tt}^\dagger)\right]^{-1}.$$

By varying the length of the disordered Fe, the corresponding resistivity was extracted [9]. In the calculations, a 48×48 **k**-point mesh was used to sample the two-dimensional Brillouin zone of the supercell, which was verified to yield well-converged results. For each length of disordered Fe, 20 random configurations of lattice disorder were included. Magnons also mediate electronic transitions with finite momentum transfer. However, the applied field suppresses magnon excitations, and we therefore neglect them in the transport calculations.

In addition to the result shown in the main text with current along [100] [Fig. 4(b)], we also calculated $\Delta\rho_2$ as a function of resistivity with current along [110], as shown in Fig. S8(a). Similar to Fig. 4(b) in the main text, the sign reversal of $\Delta\rho_2$ occurs at $\rho_{xx} \approx 3\ \mu\Omega$ cm, corresponding to ~150 K [11], consistent with experimental results of thick samples. It indicates that the sign change of $\Delta\rho_2$ in Fe induced by phonon scattering is independent of the current direction, in agreement with our experiment.

We further examined the minimal scattering model for current along [110]. Here, we used $121 \times 121 \times 121$ k-grid in the bcc Brillouin zone and a fine $\bm{k}'$-grid of $15 \times 15 \times 15$ for $q = |\bm{k}' - \bm{k}| < 0.1\bm{k}_{\text{BZ}}$ and $30 \times 30 \times 30$ for $q \geq 0.1\bm{k}_{\text{BZ}}$. The calculated $\Delta\rho_2/\rho_0$ is plotted in Fig. S8(b) as a function of the cutoff momentum transfer q, where a sign reversal is observed at a critical q value of approximately 1/20 of the Brillouin zone. These results further corroborate that scattering processes involving large momentum transfers, activated by phonon disorder, is the primary cause of the sign reversal in Fe and is independent of the current direction.

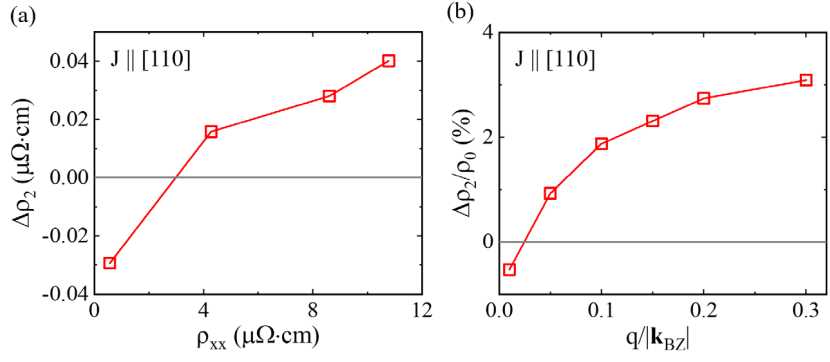

Fig. S8. (a) Calculated $\Delta\rho_2$ for current along [110], showing a sign reversal from negative to positive with increasing resistivity, using the fully quantum-mechanical transport calculation. (b) Minimal scattering-model calculation of $\Delta\rho_2$ for current along [110], confirming the same sign-change behavior.

**X. Microscopic analysis of current-direction dependence in AMR**

When we describe AMR with both an in-plane current and an in-plane magnetic field for a C$_{4v}$ system using the phenomenological model, the 4n-fold Fourier components are found to be independent of the current direction $\phi_J$. In contrast, the (4n−2)-fold components depend on $\phi_J$ with distinct coefficients $\Delta\rho_{4n-2}$ and $\Delta\rho^*_{4n-2}$ for current along [100] and [110] directions, respectively; see Eq. (S9) and Eq. (3) in the main text. To elucidate such dependence, we perform a microscopic analysis of AMR components under the C$_{4v}$ symmetry.

To be specific, we define the current direction $\boldsymbol{J} = (\cos\phi_J, \sin\phi_J, 0)$ and the magnetization direction $\mathbf{m} = (\cos\phi_{mc}, \sin\phi_{mc}, 0)$. Within the Boltzmann transport formalism, the longitudinal conductivity is given by the sum over all states at the Fermi level, denoted by $|n\boldsymbol{k}_F\rangle$,

$$\sigma(\phi_J, \phi_{mc}) = \frac{e^2}{V} \sum_{n\boldsymbol{k}_F} \left[\boldsymbol{v}_{n\boldsymbol{k}_F}(\phi_{mc}) \cdot \boldsymbol{J}\right]^2 \tau_{n\boldsymbol{k}_F}(\phi_{mc})$$

$$= \frac{e^2}{V} \sum_{n\boldsymbol{k}_F} \left[v^x_{n\boldsymbol{k}_F}(\phi_{mc}) \cos\phi_J + v^y_{n\boldsymbol{k}_F}(\phi_{mc}) \sin\phi_J\right]^2 \tau_{n\boldsymbol{k}_F}(\phi_{mc}), \quad (S12)$$

where the summation of Fermi velocity $\boldsymbol{v}_{n\boldsymbol{k}_F}$, being intrinsic properties of the Bloch state, depend only on the magnetization direction $\mathbf{m}$. The relaxation time $\tau_{n\boldsymbol{k}}$, which represents the relaxation time of the eigenstate $|n\bm{k}_F\rangle$, is also independent of the external electric field direction.

To separate the 4n and (4n−2) Fourier components with respect to the magnetization angle $\phi_{mc}$, we expand $\sigma(\phi_J, \phi_{mc})$ as a Fourier series in $\phi_{mc}$ with expansion coefficients $\sigma_i(\phi_J)$ ($i = 0, 2, 4, 6 \ldots$):

$$\sigma(\phi_J, \phi_{mc}) = \sigma_0(\phi_J) + \sigma_2(\phi_J)\cos 2\phi_{mc} + \sigma_4(\phi_J)\cos 4\phi_{mc} + \sigma_6(\phi_J)\cos 6\phi_{mc} + \cdots \tag{S13}$$

The contributions from 4n and (4n−2) terms can be rigorously isolated via the following symmetric and antisymmetric combinations:

$$\sum_n \sigma_{4n}(\phi_J)\cos(4n\phi_{mc}) = \frac{1}{2}\left[\sigma(\phi_J, \phi_{mc}) + \sigma\left(\phi_J, \phi_{mc} + \frac{\pi}{2}\right)\right] \tag{S14}$$

$$\sum_n \sigma_{4n-2}(\phi_J)\cos((4n-2)\phi_{mc}) = \frac{1}{2}\left[\sigma(\phi_J, \phi_{mc}) - \sigma\left(\phi_J, \phi_{mc} + \frac{\pi}{2}\right)\right] \tag{S15}$$

For an initial configuration $[\phi_J, \phi_{mc}]$ shown in Fig. S9(a) with the $C_{4v}$ symmetry, a 90° counterclockwise rotation of **m** with a fixed current, i.e., $\left[\phi_J, \phi_{mc} + \frac{\pi}{2}\right]$ in Fig. S9(b), is equivalent to a 90° clockwise rotation of the current direction under a fixed **m**, i.e., $\left[\phi_J - \frac{\pi}{2}, \phi_{mc}\right]$, as illustrated schematically in Fig. S9(c).

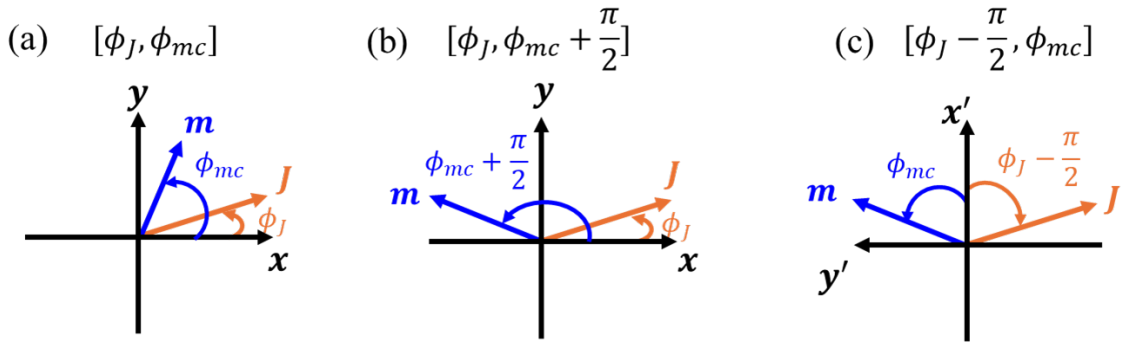

Fig. S9. Symmetry analysis of magnetization and current directions in a $C_{4v}$ system. (a) Magnetization **m** (blue arrow) and current **J** (orange arrow) directions are parameterized by azimuthal angles $\phi_{mc}$ and $\phi_J$ relative to the crystallographic x-axis, denoted as $[\phi_J, \phi_{mc}]$. (b) **m** is rotated counterclockwise by yielding a new configuration $[\phi_J, \phi_{mc} + \pi/2]$. (c) **J** is rotated clockwise by $\pi/2$, $[\phi_J - \pi/2, \phi_{mc}]$.

Note that the coordinate frame is rotated counterclockwise by π/2 about the z-axis.

Then we substitute Eq. (S12) into Eqs. (S14) and (S15) and replace $\sigma\left(\phi_J, \phi_{mc} + \frac{\pi}{2}\right)$ by $\sigma\left(\phi_J - \frac{\pi}{2}, \phi_{mc}\right)$, yielding the microscopic expressions for the summation over all 4n and (4n−2) components respectively:

$$\sum_n \sigma_{4n}(\phi_J)\cos(4n\phi_{mc}) = \frac{1}{2}\sum_{nk_F}\left\{\left[\left(v^x_{nk_F}(\phi_{mc})\right)^2 + \left(v^y_{nk_F}(\phi_{mc})\right)^2\right]\tau_{nk_F}(\phi_{mc})\right\}$$

$$\sum_n \sigma_{4n-2}(\phi_J)\cos((4n-2)\phi_{mc}) = \frac{1}{2}\sum_{nk_F}\left\{\begin{array}{l}\left[\left(v^x_{nk_F}(\phi_{mc})\right)^2 - \left(v^y_{nk_F}(\phi_{mc})\right)^2\right]\tau_{nk_F}(\phi_{mc})\cos 2\phi_J \\ +2v^x_{nk_F}(\phi_{mc})v^y_{nk_F}(\phi_{mc})\tau_{nk_F}(\phi_{mc})\sin 2\phi_J\end{array}\right\}$$

According to the orthogonality of Fourier components, the $\phi_J$ dependence of AMR coefficients $\sigma_{4n}$ vanishes for each $n$. In the same manner, each $\sigma_{4n-2}(\phi_J)$ has the twofold angular dependence on $\phi_J$.

## XI. Angular dependence of electronic velocity

In the presence of spin-orbit interaction, the electronic structure in a ferromagnetic material depends on its magnetization direction, resulting in the angular dependence of the electronic velocity near the Fermi energy. The velocity of the Bloch state |n**k**> is explicitly calculated using the standard Wannier interpolation [12], i.e. $v_{x,nk}(\phi_{mc}) = \frac{1}{\hbar}\frac{d\epsilon_{nk}(\phi_{mc})}{dk_x}$. Considering the magnetization-orientation dependence, we can expand $v_{x,nk}(\phi_{mc})$ as

$$v_{x,nk}(\phi_{mc}) = c_0^v + c_2^v \cos 2\phi_{mc} + c_4^v \cos 4\phi_{mc} + c_6^v \cos 6\phi_{mc} + \cdots. \quad (S16)$$

Figure S10(a) shows the band structure of bcc Fe along [100] near the Fermi level, where the velocities of the Bloch states in a piece of the $\Delta_5$ band (marked by magenta) are explicitly calculated. As an example, the calculated velocity of a particular Bloch state marked by the green dot in Fig. S10(a) is plotted as a function of $\phi_{mc}$ in Fig. S10(b). A fitting using Eq. (S16) up to the six-fold can well describe the calculated velocity, yielding the dominant $c_2^v$ and smaller $c_4^v$ and $c_6^v$. For the piece of the $\Delta_5$ band, the calculated two-fold, four-fold and six-fold components are plotted in Fig. S10(c). The angular dependence of velocity is dominated by the two-fold component and the four-fold component is relatively smaller. The six-fold component is very weak compared with the two-fold and four-fold ones.

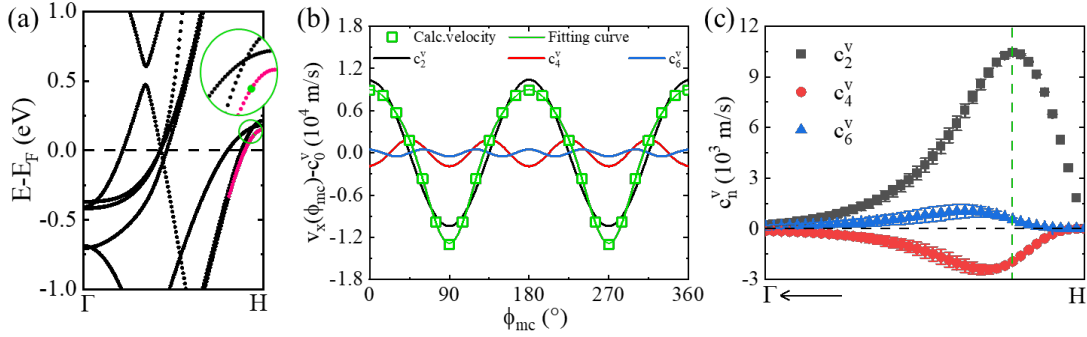

Fig. S10. (a) Band structure of bcc Fe along [100] near the Fermi level. (b) Calculated electron velocity $v_{x,n\mathbf{k}}$ for the Bloch state marked by the green dot in (a). The green curve is fitted up to the six-fold, while the two-fold, four-fold and six-fold components are plotted by the black, red and blue curves. (c) Calculated angular dependent components in the velocity $v_{x,n\mathbf{k}}$ of the Bloch states for a $\Delta_5$ band of bcc Fe (marked by magenta).

## XII. Absolute longitudinal resistivity as a function of temperature

To enable a quantitative comparison of the anisotropic magnetoresistance (AMR) amplitudes reported in this work with previous studies, we present here the absolute longitudinal resistivity $\rho_{xx}$ of epitaxial Fe(001) films as a function of temperature and film thickness.

Figure S11 shows the temperature dependence of $\rho_{xx}$ measured for Fe(001) films with thicknesses of 5.8, 9.6, 19.2, 38.4, and 78.6 nm. For all samples, the resistivity decreases monotonically upon cooling and approaches a nearly temperature-independent value below approximately 20 K, indicating that phonon scattering is strongly suppressed at low temperatures and that the residual resistivity is dominated by static disorder and interface or surface scattering.

As a representative example of film quality, the 78.6-nm-thick Fe film exhibits a resistivity of $\rho_{xx} = 10.56\ \mu\Omega\cdot\text{cm}$ at 300 K and $\rho_{xx} = 0.285\ \mu\Omega\cdot\text{cm}$ at 5 K, corresponding to a residual resistivity ratio (RRR) of approximately 37. This large RRR reflects the high crystalline quality and low defect density of the thick epitaxial films. In contrast, thinner films exhibit higher residual resistivity and smaller RRR values, consistent with the enhancement of surface and interface scattering.

These absolute resistivity data provide an essential reference for evaluating the magnitude of the AMR coefficients reported in the main text and allow direct comparison with previous experimental studies on Fe and other cubic ferromagnets.

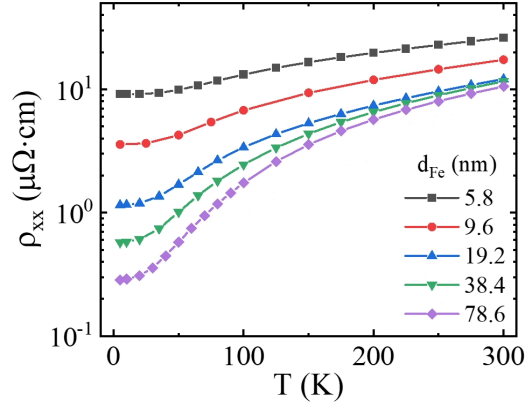

Fig. S11. Temperature dependence of the longitudinal resistivity $\rho_{xx}$ for Fe(001) films with thicknesses of 5.8, 9.6, 19.2, 38.4, and 78.6 nm.

## XIII. Effect of Multi-Cycle Rotation on FFT Analysis of AMR and PHE

In angular-dependent magnetotransport measurements, experimental data obtained in different rotation cycles are not strictly identical due to unavoidable measurement noise. This noise includes not only high-frequency electronic noise but also low-frequency resistivity fluctuations induced by slow temperature variations in the low-temperature measurement environment. As a result, performing multiple rotation cycles provides an effective way to average out non-periodic noise components and thereby improve the signal-to-noise ratio of the angular-dependent resistivity.

In the present measurements, the rotation rate was already sufficiently slow to ensure an angular sampling interval of the magnetization angle $\phi_m$ below 1°. Further reducing the rotation rate for a single-cycle measurement does not significantly improve data quality, because it primarily suppresses high-frequency noise while remaining ineffective against low-frequency fluctuations. By contrast, averaging over multiple rotation cycles efficiently suppresses both high- and low-frequency non-periodic noise.

To explicitly demonstrate the benefit of multi-cycle rotation, we compared the angular harmonics of anisotropic magnetoresistance (AMR) and planar Hall effect (PHE) extracted by fast Fourier transform (FFT) from ten-cycle rotations with those obtained from a single rotation cycle. Figures S12(a) and S12(b) show the FFT spectra of ten consecutive in-plane rotations for current directions J∥[100] and J∥[110], respectively. For comparison, Figs. S12(c) and S12(d) show the corresponding FFT spectra obtained from a single in-plane rotation.

Because the total angular range is extended by a factor of ten, the FFT of ten-cycle rotations achieves a frequency resolution of 1/10-fold, compared to 1-fold for a single-cycle rotation, and exhibits a noise floor that is at least one order of magnitude lower. This improvement enables reliable resolution of high-order angular harmonics that are otherwise obscured by noise in single-cycle measurements.

To further compare FFT analysis with conventional fitting, we also analyzed the same data using even-order cosine or sine series truncated at the 18th order. The fitting results are shown as column plots in Fig. S12(a-d), overlaid with the FFT spectra (curves). The fitted harmonic amplitudes are fully consistent with those obtained from FFT, demonstrating the equivalence of the two methods in extracting angular harmonics. Importantly, the fitting error bars—shown on top of the columns—are significantly smaller for the ten-cycle data than for the single-cycle data, particularly for harmonics above the 10th order, further confirming the enhanced signal quality achieved by multi-cycle rotation.

Although FFT and Fourier-series fitting are formally equivalent, FFT offers practical advantages for identifying high-order AMR and PHE harmonics. In particular, FFT does not rely on a predefined fitting function and provides a direct visualization of the noise floor associated with non-periodic signals. By comparing the amplitude of a given harmonic with the noise background, the reliability of high-order components can be assessed in a transparent and model-independent manner.

Finally, we note that the imperfect reciprocity between AMR and PHE coefficients listed in Tables S1 and S2 does not originate from measurement noise. The typical fitting uncertainty for ten-cycle data is on the order of 0.002 nΩ·cm, which is much smaller than the observed deviations. Instead, these differences arise from device-to-device variations, as measurements for J∥[100] and J∥[110] were performed on different Hall bar devices that may exhibit slight differences in geometry or crystalline quality.

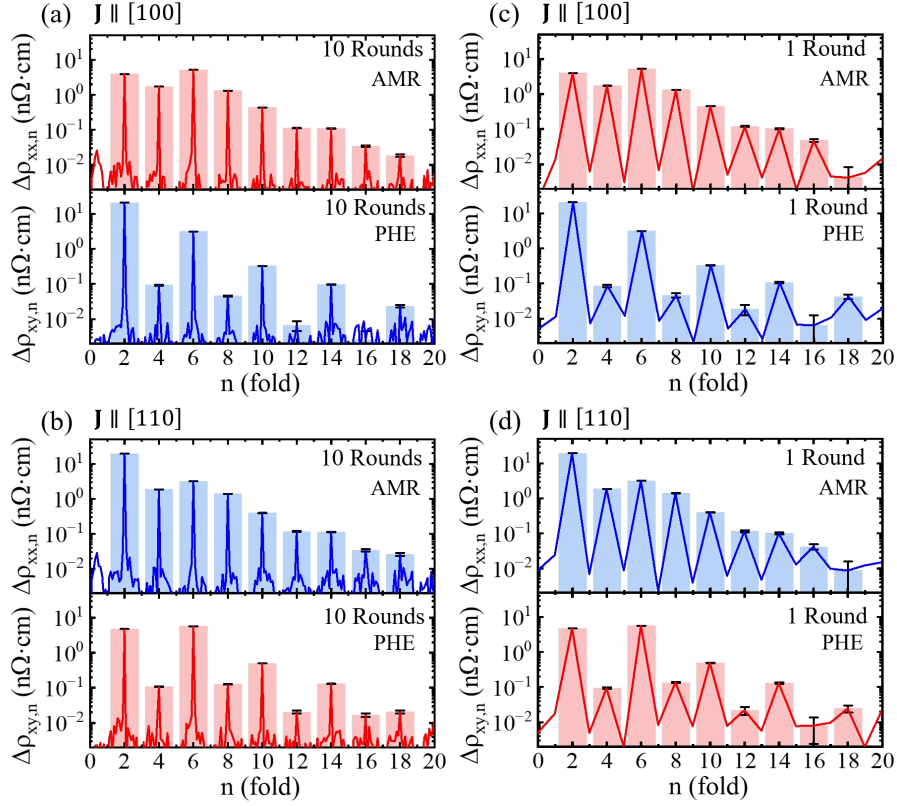

Fig. S12. Comparison of angular-harmonic analysis obtained from ten-cycle and single-cycle rotation measurements. All data were taken in the 5.8-nm-thick sample at a temperature of 5 K. (a,b) FFT spectra of AMR and PHE for ten consecutive in-plane rotations with current along (a) J∥[100] and (b) J∥[110]. (c,d) Corresponding FFT spectra obtained from a single in-plane rotation for (c) J∥[100] and (d) J∥[110]. Column plots in (a-d) show the amplitudes extracted from Fourier-series fitting using even-order cosine or sine terms truncated at the 18th order; the columns overlap with the FFT spectra (curves), demonstrating the equivalence of the two analysis methods.

# References

[1] A. L. Ravensburg, M. Werwiński, J. Rychły-Gruszecka, J. Snarski-Adamski, A. Elsukova, P. O. Å. Persson, J. Rusz, R. Brucas, B. Hjörvarsson, P. Svedlindh, G. K. Pálsson, and V. Kapaklis, Boundary-induced phase in epitaxial iron layers. Phys. Rev. Mater. **8**, L081401 (2024).

[2] H. Chen, Z. Cheng, Y. Feng, H. Xu, T. Wu, C. Chen, Y. Chen, Z. Yuan, and Y. Wu, Anisotropic galvanomagnetic effects in single-crystal Fe(001) films elucidated by a phenomenological theory, Phys. Rev. B **111**, 014437 (2025).

[3] P. Granberg, P. Isberg, T. Baier, B. Hjörvarsson, and P. Nordblad, Anisotropic behaviour of the magnetoresistance in single crystalline iron films, J. Magn. Magn. Mater. **195**, 1 (1999).

[4] R. P. van Gorkom, J. Caro, T. M. Klapwijk, and S. Radelaar, Temperature and angular dependence of the anisotropic magnetoresistance in epitaxial Fe films, Phys. Rev. B **63**, 134432 (2001).

[5] T. McGuire and R. Potter, Anisotropic magnetoresistance in ferromagnetic 3d alloys, IEEE Trans. Magn. **11**, 1018 (1975).

[6] W. Döring, Die abhängigkeit des widerstandes von nickelkristallen von der richtung der spontanen magnetisierung, Ann. Phys. **424**, 259 (1938).

[7] R. R. Birss, *Symmetry and magnetism* (North-Holland, Amsterdam, 1964).

[8] F. L. Zeng, Z. Y. Ren, Y. Li, J. Y. Zeng, M. W. Jia, J. Miao, A. Hoffmann, W. Zhang, Y. Z. Wu, and Z. Yuan, Intrinsic mechanism for anisotropic magnetoresistance and experimental confirmation in CoxFe1-x single-crystal films, Phys. Rev. Lett. **125**, 097201 (2020).

[9] A. A. Starikov, Y. Liu, Z. Yuan, and P. J. Kelly, Calculating the transport properties of magnetic materials from first principles including thermal and alloy disorder, noncollinearity, and spin-orbit coupling, Phys. Rev. B **97**, 214415 (2018).

[10] Y. Liu, A. A. Starikov, Z. Yuan, and Paul J. Kelly. First-principles calculations of magnetization relaxation in pure Fe, Co, and Ni with frozen thermal lattice disorder. Phys. Rev. B **84**, 014412 (2011).

[11] C. Y. Ho, M. W. Ackerman, K. Y. Wu, T. N. Havill, R. H. Bogaard, R. A. Matula, S. G. Oh and H. M. James. Electrical resistivity of ten selected binary alloy systems. J. Phys. Chem. Ref. Data **12**, 183 (1983).

[12] J. R. Yates, X. Wang, D. Vanderbilt, and I. Souza. Spectral and Fermi surface properties from Wannier interpolation. Phys. Rev. B **75**, 195121 (2007).